\documentclass[10pt,twocolumn,english,aps,showpacs,preprintnumbers,nofootinbib,prd,superscriptaddress,groupedaddress]{revtex4-2}
\usepackage{beramono}
\usepackage[T1]{fontenc}
\usepackage[latin9]{inputenc}
\usepackage{color}
\usepackage{babel}
\usepackage{verbatim}
\usepackage{mathtools}
\usepackage{amsmath}
\usepackage{amsthm}
\usepackage{amssymb}
\usepackage{graphicx}
\usepackage{wasysym}
\usepackage{esint}
\usepackage[unicode=true,
 bookmarks=false,
 breaklinks=false,pdfborder={0 0 1},backref=false,colorlinks=true]
 {hyperref}
\hypersetup{
 citecolor=darkblue,linkcolor=darkblue}

\makeatletter
\usepackage{babel}
\usepackage{amsthm}
\usepackage{amsfonts}
\usepackage{epsfig}
\usepackage{times}

\usepackage{epstopdf}
\usepackage{aas_macros}
\usepackage{tensor}
\usepackage{bm}
\usepackage{url}\usepackage{wasysym}

\@ifundefined{textcolor}{}{%
 \definecolor{BLACK}{gray}{0}
 \definecolor{WHITE}{gray}{1}
 \definecolor{RED}{rgb}{1,0,0}
 \definecolor{GREEN}{rgb}{0,1,0}
 \definecolor{BLUE}{rgb}{0,0,1}
 \definecolor{CYAN}{cmyk}{1,0,0,0}
 \definecolor{MAGENTA}{cmyk}{0,1,0,0}
 \definecolor{YELLOW}{cmyk}{0,0,1,0}
}
\definecolor{darkblue}{rgb}{0,0,0.5}
\usepackage{soul}

\usepackage{aecompl}

\makeatother

\begin{document}
\title{Relativistic effects cannot explain galactic dynamics}
\author{L.~Filipe~O.~Costa}
\email{lfilipecosta@tecnico.ulisboa.pt}

\affiliation{CAMGSD, Departamento de Matemática, Instituto Superior Técnico, 1049-001
Lisboa, Portugal}
\author{José~Natário}
\email{jnatar@math.ist.utl.pt}

\affiliation{CAMGSD, Departamento de Matemática, Instituto Superior Técnico, 1049-001
Lisboa, Portugal}
\date{\today}
\begin{abstract}
It has been suggested in recent literature that nonlinear and/or gravitomagnetic
general relativistic effects can play a leading role in galactic dynamics,
partially or totally replacing dark matter. Using the 1+3 ``quasi-Maxwell''
formalism, we show, on general grounds, such hypothesis to be impossible. 
\end{abstract}
\maketitle
\tableofcontents{}

\section{Introduction}

In a recent number of works, based both on linearized theory \cite{RuggieroGalatic,AstesianoRuggieroPRDII}
and exact models \cite{CooperstockTieu,CarrickCooperstock,BG,Crosta2018,RuggieroBG},
it has been asserted that general relativistic effects (namely, gravitomagnetic
ones) can have a significant impact \cite{RuggieroBG,RuggieroGalatic,AstesianoRuggieroPRDII,BG},
or even totally account for the galactic flat rotation curves \cite{Crosta2018,CooperstockTieu,CarrickCooperstock}.
In the framework of a weak field slow motion approximation, this has
been shown to be impossible, and such claims addressed, in Refs. \cite{Ciotti:2022inn,LasenbyHobsonBarker,GlampedakisJones_Pitfalls}.
It has however been argued \cite{CooperstockTieu,CarrickCooperstock,BG,RuggieroBG}
that it is still possible in the exact theory, due to nonlinear effects
not captured in linearized theory (and not manifest ``locally''
\cite{CooperstockTieu,CarrickCooperstock,BG}). In support of these
claims, the Balasin-Grumiller (BG) model \cite{BG,Crosta2018,RuggieroBG},
or variants of it \cite{CooperstockTieu,CarrickCooperstock,RuggieroBG},
have been used. Such model has been recently debunked in our Ref.
\cite{Costa_e_al_Frames}, where it was shown to actually consist
of a static dust, held in place by unphysical singularities along
the symmetry axis, and that what actually rotates in the model are
the ill chosen reference observers. In the present paper we address
the general problem: assuming only stationarity of the solution, we
show that (i) gravitomagnetism cannot be a relevant driving force
of galactic dynamics; (ii) nonlinear general relativistic effects
cannot account for any amount of the needed dark matter. 

\emph{Notation and conventions.---} \textcolor{black}{Greek letters
$\alpha,\beta,\gamma,...$ denote 4D spacetime indices, running 0--3;
Roman letters $i,j,k,...$ are spatial indices, running 1--3}; $\epsilon_{\alpha\beta\gamma\delta}\equiv\sqrt{-g}[\alpha\beta\gamma\delta]$
is the 4D Levi-Civita tensor, with the orientation $[1230]=1$; $\epsilon_{ijk}\equiv\sqrt{h}[ijk]$\textcolor{black}{{}
is the} Levi-Civita tensor in a 3D Riemannian manifold $\Sigma$ of
metric $h_{ij}$. Arrow notation $\vec{X}$ denotes spatial 3-vectors
on $\Sigma$, with components $X^{i}$. Indices of spatial tensors
on $\Sigma$ are raised and lowered with $h_{ij}$: $X_{i}=h_{ij}X^{j}$.
The basis vector corresponding to a coordinate $\phi$ is denoted
by $\partial_{\phi}\equiv\partial/\partial\phi$, and its $\alpha$-component
by $\partial_{\phi}^{\alpha}\equiv\delta_{\phi}^{\alpha}$.

\section{Equations of motion for massive particles and light in a stationary
field}

The line element $ds^{2}=g_{\alpha\beta}dx^{\alpha}dx^{\beta}$ of
a stationary spacetime can generically be written as 
\begin{equation}
ds^{2}=-e^{2\Phi}(dt-\mathcal{A}_{i}dx^{i})^{2}+h_{ij}dx^{i}dx^{j}\ ,\label{eq:StatMetric}
\end{equation}
where $e^{2\Phi}=-g_{00}$, $\Phi\equiv\Phi(x^{j})$, $\mathcal{A}_{i}\equiv\mathcal{A}_{i}(x^{j})=-g_{0i}/g_{00}$,
and $h_{ij}(x^{k})=g_{ij}+e^{2\Phi}\mathcal{A}_{i}\mathcal{A}_{j}$.
We are interested in the equations of motion with respect to the coordinate
system in \eqref{eq:StatMetric}; it is thus useful to refer to quantities
as measured by the observers at rest in such coordinates (``static''
observers). These observers have 4-velocity $u^{\alpha}=e^{-\Phi}\partial_{t}^{\alpha}\equiv e^{-\Phi}\delta_{0}^{\alpha}$,
and their worldlines are tangent to the timelike Killing vector field
$\partial_{t}$. The quotient of the spacetime by their worldlines
yields a 3D Riemannian manifold $\Sigma$ with metric $h_{ij}$, which
these observers regard as the ``spatial metric'', since $dl=\sqrt{h_{ij}dx^{i}dx^{j}}$
yields the infinitesimal spatial distance between them as measured
by Einstein's light signaling procedure (e.g., by radar) \cite{LandauLifshitz}.
The 3D metric $h_{ij}$ is identified in spacetime with the \emph{space
projector} with respect to $u^{\alpha}$, $h_{\alpha\beta}\equiv g_{\alpha\beta}+u_{\alpha}u_{\beta}$
(observe that $h_{0\alpha}=0$).

As discussed in detail in \cite{Costa_e_al_Frames}, when the congruence
of static observers is asymptotically inertial {[}which is the case
when their acceleration and vorticity, Eq. (7) below, asymptotically
vanish{]}, the coordinate system in \eqref{eq:StatMetric} has axes
anchored to inertial frames at infinity, being a generalization of
the International Astronomical Union reference system to the exact
metric (\ref{eq:StatMetric}). Such reference frame is in practice
realized setting it up fixed with respect to remote celestial objects
(namely quasars), and it is with respect to such a frame that, for
instance, galactic rotation curves are defined. 

\subsection{Geodesic equation in the \textquotedblleft quasi-Maxwell\textquotedblright{}
formalism}

Let $x^{\alpha}(\lambda)$ be the geodesic worldline of a test particle
(that can have zero mass) and $\lambda$ an affine parameter along
it. The space components of the geodesic equation $d^{2}x^{\alpha}/d\lambda^{2}=-\Gamma_{\mu\nu}^{\alpha}(dx^{\mu}/d\lambda)(dx^{\nu}/d\lambda)$
can be written as\footnote{\label{fn:Christoffel}Using the Christoffel symbols $\Gamma_{00}^{i}=-e^{2\Phi}G^{i}$,
$\Gamma_{j0}^{i}=e^{2\Phi}\mathcal{A}_{j}G^{i}-e^{\Phi}H_{\ j}^{i}/2$,
and $\Gamma_{jk}^{i}=\Gamma(h)_{jk}^{i}-e^{\Phi}\mathcal{A}_{(k}H_{j)}^{\ i}-e^{2\Phi}G^{i}\mathcal{A}_{j}\mathcal{A}_{k}$,
where $H_{ij}\equiv e^{\Phi}[\mathcal{A}_{j,i}-\mathcal{A}_{i,j}]$.}
\begin{equation}
\frac{d^{2}x^{i}}{d\lambda^{2}}+\Gamma(h)_{jk}^{i}\frac{dx^{j}}{d\lambda}\frac{dx^{k}}{d\lambda}=\nu\left[\nu\vec{G}+\frac{d\vec{x}}{d\lambda}\times\vec{H}\right]^{i}\label{eq:QMGeoGeneral}
\end{equation}
where $\nu\equiv-u_{\alpha}dx^{\alpha}/d\lambda$, 
\begin{equation}
\Gamma(h)_{jk}^{i}\equiv\frac{1}{2}h^{il}\left(h_{lj,k}+h_{lk,j}-h_{jk,l}\right)\label{eq:ChristoffelSpace}
\end{equation}
 are the Christoffel symbols of the space metric $h_{ij}$, and 
\begin{equation}
G_{i}=-\Phi_{,i}\ ;\qquad\ H^{i}=e^{\Phi}\epsilon^{ijk}\mathcal{A}_{k,j}\quad(\epsilon_{ijk}\equiv\sqrt{h}[ijk])\label{eq:GEM1forms}
\end{equation}
are fields living on the space manifold $\Sigma$, dubbed, respectively,
``gravitoelectric'' and ``gravitomagnetic'' fields, for playing
in Eq. \eqref{eq:QMGeoGeneral} roles analogous to the electric and
magnetic fields in the Lorentz force equation. For a 3-vector $\vec{X}$
on $\Sigma$, one can define the 3D covariant derivative with respect
to $h_{ij}$ 
\begin{equation}
\frac{\tilde{D}X^{i}}{d\lambda}\equiv\frac{dX^{i}}{d\lambda}+\Gamma(h)_{jk}^{i}X^{j}\frac{dx^{k}}{d\lambda}\ .\label{eq:TildaDerivative}
\end{equation}

\subsubsection{Timelike geodesics}

For timelike worldlines, one can set $\lambda=\tau$, so $dx^{\alpha}/d\tau\equiv U^{\alpha}$
is the 4-velocity, $\nu=-u_{\alpha}U^{\alpha}\equiv\gamma$ becomes
the Lorentz factor between $U^{\alpha}$ and $u^{\alpha}$, and Eq.
\eqref{eq:QMGeoGeneral} can therefore be written as \cite{LandauLifshitz,ZonozBell1998,NatarioQM2007,Analogies,Cilindros,Zonoz2019}
(cf. also \cite{ManyFaces,BiniIntrinsic}), 
\begin{equation}
\frac{\tilde{D}\vec{U}}{d\tau}=\gamma\left[\gamma\vec{G}+\vec{U}\times\vec{H}\right]\label{eq:QMGeo}
\end{equation}
where $\vec{U}$ is the 3-vector of components $(\vec{U})^{i}=U^{i}$,
tangent to the spatial curve $x^{i}(\tau)$ obtained by projecting
the geodesic onto $\Sigma$. Equation (\ref{eq:QMGeo}) describes
the acceleration of such 3D projected curve {[}since the left-hand
member of (\ref{eq:QMGeoGeneral}) is the standard 3D covariant acceleration{]}.
Its physical interpretation is that, from the point of view of the
static observers, the spatial trajectory of the test particle will
appear accelerated, as if acted upon by fictitious forces (inertial
forces), arising from the fact that the reference frame is \emph{not
inertial}. In fact, $\vec{G}$ and $\vec{H}$ are identified in spacetime,
respectively, with minus the acceleration and twice the vorticity
of the static observers: 
\begin{equation}
G^{\alpha}=-\nabla_{\mathbf{u}}u^{\alpha}\equiv-u_{\ ;\beta}^{\alpha}u^{\beta}\;;\qquad H^{\alpha}=2\omega^{\alpha}=\epsilon^{\alpha\beta\gamma\delta}u_{\gamma;\beta}u_{\delta}\;.\label{eq:GEM Fields Cov}
\end{equation}
Here $\epsilon^{\alpha\beta\gamma\delta}=[\alpha\beta\gamma\delta]/\sqrt{-g}$
is the 4D contravariant Levi-Civita tensor, and in the correspondence
with \eqref{eq:GEM1forms} it is useful to note that $-g=he^{2\Phi}$.
Decomposing $U^{\alpha}$ in terms of its projections parallel and
orthogonal to $u^{\alpha}$, 
\begin{equation}
U^{\alpha}=\gamma(u^{\alpha}+v^{\alpha})\ ;\qquad\gamma\equiv-u^{\alpha}U_{\alpha}=\frac{1}{\sqrt{1-v^{\alpha}v_{\alpha}}}\ ,\label{eq:u_u'}
\end{equation}
where $v^{\alpha}$ is the spatial velocity of the particle relative
to the static observers \cite{ManyFaces,Costa_e_al_Frames,Analogies},
Eq. \eqref{eq:QMGeo} can be rewritten as 
\begin{equation}
\frac{\tilde{D}\vec{U}}{d\tau}=\gamma^{2}\left[\vec{G}+\vec{v}\times\vec{H}\right]\ .\label{eq:QMGeov}
\end{equation}

\subsubsection{Null geodesics}

In the case of light rays (null geodesics), of tangent $k^{\alpha}=dx^{\alpha}/d\lambda$,
Eqs. \eqref{eq:QMGeoGeneral} and \eqref{eq:TildaDerivative} analogously
yield 
\begin{equation}
\frac{\tilde{D}\vec{k}}{d\lambda}=\nu\left[\nu\vec{G}+\vec{k}\times\vec{H}\right]\ ;\qquad\nu\equiv-k^{\alpha}u_{\alpha}\ .\label{eq:Geolight}
\end{equation}
An equivalent formulation in terms of the photon's null 4-momentum
is given in \cite{BiniIntrinsic}, Eqs. (10.2)--(10.3), and a Lagrangian
formulation in \cite{Perlick_Lensing_2004}. The null vector $k^{\alpha}$
decomposes, in terms of its projections parallel and orthogonal to
$u^{\alpha}$, as \cite{BolosIntrinsic}
\begin{equation}
k^{\alpha}=\nu(u^{\alpha}+v^{\alpha})\label{eq:vrelLight}
\end{equation}
where $v^{\alpha}$ is a \emph{unit} vector orthogonal to $u^{\alpha}$,
yielding the photon's spatial velocity relative to $u^{\alpha}$.
Therefore, $\vec{k}=\nu\vec{v}$, and one can re-write \eqref{eq:Geolight}
as 
\begin{equation}
\frac{\tilde{D}\vec{k}}{d\lambda}=\nu^{2}\left[\vec{G}+\vec{v}\times\vec{H}\right]\ .\label{eq:Geolightv}
\end{equation}
Equations \eqref{eq:Geolight}-\eqref{eq:Geolightv} tell us that,
just like massive particles, light rays in a stationary spacetime
behave analogously to charged particles under the action of an electric
and magnetic fields \cite{Perlick_Lensing_2004}. A difference, however,
is that the second term in the left-hand member of \eqref{eq:QMGeoGeneral},
which reads here $\Gamma(h)_{jm}^{i}k^{j}k^{m}$, is typically of
the same order magnitude as the first term in \eqref{eq:Geolight}
{[}contrary to the case of $\Gamma(h)_{jm}^{i}U^{j}U^{m}=O(v^{2})$
for slowly moving massive particles{]}.

\subsubsection{Timelike circular geodesics}

In an axistationary spacetime we have $g_{0i}dx^{i}=g_{0\phi}d\phi$.
If there is reflection symmetry about the equatorial plane \cite{SajalReflection2021},
then $g_{\alpha\beta,z}=0$ (or $g_{\alpha\beta,\theta}=0$) along
that plane; noticing that the geodesic equation can be written in
the form (see e.g. \cite{CarrollBook}) $dU_{\alpha}/d\tau=g_{\mu\nu,\alpha}U^{\mu}U^{\nu}/2$,
then $dU_{z}=0$ therein, allowing equatorial geodesic motion. For
circular equatorial geodesics, $U^{\alpha}=U^{0}(\delta_{0}^{\alpha}+\Omega_{{\rm geo}}\delta_{\phi}^{\alpha})$,
where the angular velocity $\Omega_{{\rm geo}}\equiv d\phi/dt=U^{\phi}/U^{0}$
is obtained from the $r$-component of the geodesic equation, $dU_{r}/d\tau=g_{\mu\nu,r}U^{\mu}U^{\nu}/2=0$:
\begin{align}
\Omega_{{\rm geo}\pm} & =\frac{-g_{0\phi,r}\pm\sqrt{g_{0\phi,r}^{2}-2g_{\phi\phi,r}e^{2{\rm \Phi}}G_{r}}}{g_{\phi\phi,r}}\ .\label{eq:OmegaGeo}
\end{align}
In the equatorial plane, given an arbitrary radial coordinate $r$,
through the coordinate transformation $r_{{\rm w}}(r)=e^{\Phi(r)}\sqrt{h_{\phi\phi}(r)}$,
the metric can be written in the form \eqref{eq:StatMetric} with
$h_{\phi\phi}=e^{-2\Phi}r_{{\rm w}}^{2}$. This is the case of the
Weyl canonical coordinates (cf. e.g. Eq. (19.21) of \cite{StephaniExact});
it is also the case (to the accuracy at hand) of the radial coordinate
$r=\sqrt{x^{2}+y^{2}}$ associated to the post-Newtonian (PN) coordinate
system in Eq. \eqref{eq:PNmetric} below. It follows from \eqref{eq:u_u'}
that the magnitude $\|v_{{\rm geo\pm}}^{\alpha}\|=(1-\gamma_{{\rm geo\pm}}^{-2})^{1/2}$
of the velocity of the circular geodesics relative to the static observers
$u^{\alpha}=e^{-\Phi}\delta_{0}^{\alpha}$ reads, in such coordinates,
for $G_{r_{{\rm w}}}\ne0$,
\begin{equation}
\|v_{{\rm geo\pm}}^{\alpha}\|=\frac{2r_{{\rm w}}|G_{r_{{\rm w}}}|}{|\sqrt{-g(H^{z})^{2}-4r_{{\rm w}}G_{r_{{\rm w}}}(1+r_{{\rm w}}G_{r_{{\rm w}}})}\pm\sqrt{-g}H^{z}|}\label{eq:vgeo}
\end{equation}
where we have noted, from \eqref{eq:GEM1forms}, that $\mathcal{A}_{\phi,r}=e^{-\Phi}\epsilon_{r\phi i}H^{i}=e^{-2\Phi}\sqrt{-g}H^{z}$.

\subsection{Field equations}

The fields $\vec{G}$ and $\vec{H}$ that govern the geodesic equation
\eqref{eq:QMGeoGeneral} obey \cite{Analogies} 
\begin{align}
 & \tilde{\nabla}\cdot\vec{G}=-4\pi(2\rho+T_{\ \alpha}^{\alpha})+{\vec{G}}^{2}+\frac{1}{2}{\vec{H}}^{2}\ ;\qquad\tilde{\nabla}\times\vec{G}=\ 0\ ;\label{eq:GFieldEq}\\
 & \tilde{\nabla}\cdot\vec{H}=-\vec{G}\cdot\vec{H}\;;\qquad\tilde{\nabla}\times\vec{H}=-16\pi\vec{J}+2\vec{G}\times\vec{H}\ ,\label{eq:HFieldEq}
\end{align}
where $\rho\equiv T^{\alpha\beta}u_{\alpha}u_{\beta}$ and $J^{\alpha}\equiv-T^{\alpha\beta}u_{\beta}$
are, respectively, the mass-energy density and current 4-vector as
measured by the static observers of 4-velocity $u^{\alpha}=e^{-\Phi}\delta_{0}^{\alpha}$.
Here $\tilde{\nabla}$ denotes covariant differentiation with respect
to the spatial metric $h_{ij}$, with Christoffel symbols \eqref{eq:ChristoffelSpace}.
The equations for $\tilde{\nabla}\cdot\vec{G}$ and $\tilde{\nabla}\times\vec{H}$
are, respectively, the time-time and time-space projections, with
respect to $u^{\alpha}$, of the Einstein field equations $R_{\alpha\beta}=8\pi(T_{\alpha\beta}^{\ }-\frac{1}{2}g_{\alpha\beta}^{\ }T_{\ \gamma}^{\gamma})$;
the equations for $\tilde{\nabla}\cdot\vec{H}$ and $\tilde{\nabla}\times\vec{G}$
follow from \eqref{eq:GEM1forms}.

\section{Gravitomagnetism cannot be the culprit---gravitational lensing\label{subsec:Lensing}}

\begin{figure}
\includegraphics[width=1\columnwidth]{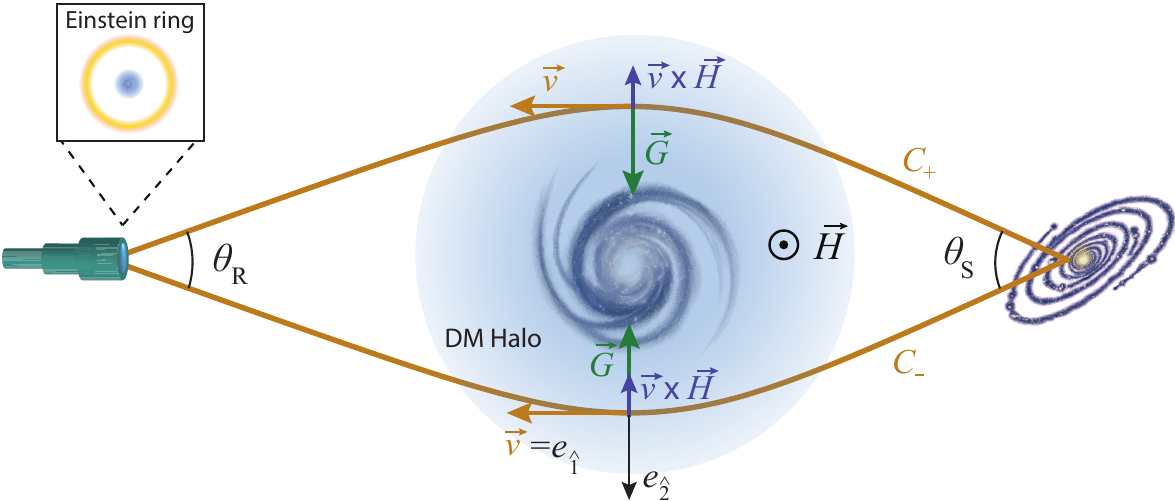}\caption{\label{fig:EinsteinRing}Gravitational lensing by galactic dark matter
halos. When the observer, foreground, and emitting galaxies are aligned,
an Einstein ring forms. Nearly perfect Einstein rings have been detected,
such as in the system B1938+666, or the \textquotedblleft Cosmic Horseshoe\textquotedblright{}
J1148+1930. The majority of the lensing effect is estimated to come
from dark matter. The gravitomagnetic field cannot mimic such effects,
since the inertial force $\vec{v}\times\vec{H}$ does not contribute
to make light rays converge in such setting; e.g., in the equatorial
plane, it deflects photons on both sides of the body in the same direction
(the effect being anyway negligible for any realistic galactic model).}

\end{figure}

Alongside the galactic rotation curves, there is a set of other effects
consistently indicating hypothetical dark matter (DM) halos around
galaxies. Any viable alternative to DM must explain these effects.
One of the most compelling of such indications is the observed galactic
gravitational lensing, which cannot be accounted for by visible baryonic
matter alone. It is well known that, when the emitting object (light
source), the foreground galaxy (lens), and the observer are nearly
aligned (see Fig \ref{fig:EinsteinRing}), roughly circular rings
(Einstein rings \cite{EinsteinRing,Pinochet_Ring,Petters_2001singularity,schneider_Ehlers_Falco_lensing,Virbhadra_Lensing2008,SalucciReview,Kochanek_et_al_Rings})
are formed, as in the case of the system B1938+666 \cite{King_first_Einstein_Ring,Lagattuta_Einstein_ring,Petters_2001singularity}
(the first detected complete Einstein ring), or the ``Cosmic Horseshoe''
J1148+1930 \cite{Belokurov_Horseshoe_2007,Spiniello_horseshoe_2011,Bellagamba_horseshoe_2011,Schuldt_et_al_horseshoe_2019,Cheng_et_al_horseshoe_2019}.
The majority of the mass causing such lensing effect is estimated
to consist of dark matter \cite{Spiniello_horseshoe_2011,Bellagamba_horseshoe_2011,Schuldt_et_al_horseshoe_2019,Cheng_et_al_horseshoe_2019,Lagattuta_Einstein_ring},
and is consistent with the halos' shape being roughly spherical or
moderately deformed (namely prolate \cite{Schuldt_et_al_horseshoe_2019},
as seems to be more common \textcolor{blue}{\cite{Durkalec}}). Anomalies
in such rings have actually been recently proposed \cite{Amruth_ring_Anomalies_2023}
as a mean of determining the nature of dark matter, namely distinguishing
between weakly interacting massive particles and axions. 

\begin{figure}
\includegraphics[width=0.8\columnwidth]{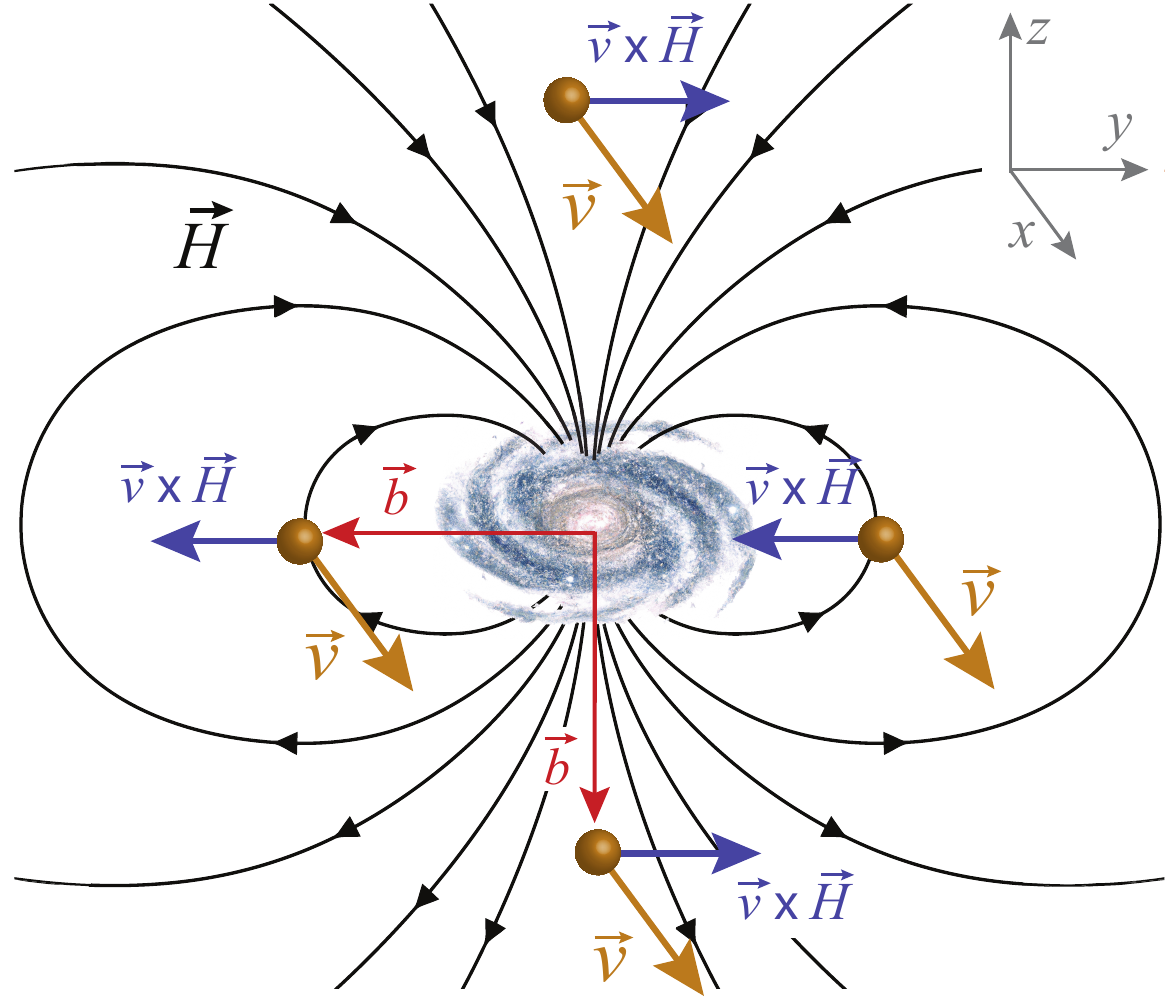}

\caption{\label{fig:DipoleDeflection}Gravitomagnetic field $\vec{H}$ produced
by a galactic disk. At large distances, it is dipole-like, deflecting
photons with impact parameter $\vec{b}$ orthogonal to the equatorial
plane in a direction opposite to those for which $\vec{b}$ lies in
the equatorial plane (in both cases, photons through opposite sides
of the body being deflected in the same direction); this produces
no convergence of light rays along the line of sight of observers
aligned with the light source and the foreground galaxy.}
\end{figure}

These effects could not take place if the gravitomagnetic field $\vec{H}$
were the main driver of galactic dynamics. For a rotating body, under
the assumptions of stationarity and reflection symmetry about the
equatorial plane, $\vec{H}$ is axisymmetric, orthogonal to the equatorial
plane, and pointing in the same direction along the whole plane. It
follows from Eq. \eqref{eq:Geolightv} that the gravitomagnetic ``force''
$\vec{v}\times\vec{H}$ on light rays passing through opposite sides
of the body points in the same direction, thus not contributing to
produce symmetric convergence of light rays along the line of sight
of observers aligned with the light source and the foreground galaxy
(setting illustrated in Fig. \ref{fig:EinsteinRing}). This can be
seen also using the Gauss-Bonnet theorem \textcolor{blue}{\cite{Ono_GM_lensing_2017,BG_lensing_2022,Ishihara_et_al_lensing2016,Gibbons_Werner_2008,Jusufi_lensing_2018,HallaPerlick2020,Sanchez_2023}}:
the light ray trajectories $C_{\pm}$ illustrated in Fig. \ref{fig:EinsteinRing}
are the projections of the photon's null geodesics $C_{\pm}^{(4)}$
onto the space manifold $\Sigma$, $\pi_{\Sigma}C_{\pm}^{(4)}=C_{\pm}$.
Let $\mathcal{S}$ be an oriented 2-surface on $\Sigma$ bounded by
$C_{\pm}$: $\partial\mathcal{S}=C_{+}\cup(-C_{-})$, where $-C_{-}$
is the curve with the opposite orientation of $C_{-}$. Then
\begin{equation}
\theta_{{\rm R}}=\iint_{\mathcal{S}}Kd\mathcal{S}+\int_{C_{+}}\kappa_{{\rm g}}d\lambda-\int_{C_{-}}\kappa_{{\rm g}}d\lambda-\theta_{{\rm S}}-2\pi[\chi(\mathcal{S})-1]\ ,\label{eq:ConvergenceGaussBonnet}
\end{equation}
where $K$ is the Gaussian curvature of $\mathcal{S}$ and $\chi(\mathcal{S})$
its Euler characteristic {[}if simply connected, namely in the absence
of singularities, $\chi(\mathcal{S})=1${]}, and $\kappa_{{\rm g}}$
is the geodesic curvature \cite{klingenberg_book,godinhoNatario,Ono_GM_lensing_2017}
of a curve $C$ of tangent vector $\dot{C}(\lambda)$. Considering
a positively oriented orthonormal frame $\{e_{\hat{1}},e_{\hat{2}}\}$
on $\mathcal{S}$ along $C$ such that $e_{\hat{1}}=\dot{C}/\|\dot{C}\|$
(see Fig. \ref{fig:EinsteinRing}), the geodesic curvature of $C$
is given by \cite{klingenberg_book} 
\[
\kappa_{{\rm g}}=\frac{1}{\|\dot{C}\|}\langle\tilde{\nabla}_{\dot{C}}e_{\hat{1}},e_{\hat{2}}\rangle=\frac{\langle\tilde{\nabla}_{\dot{C}}\dot{C},e_{\hat{2}}\rangle}{\|\dot{C}\|^{2}}\ ,
\]
where $\tilde{\nabla}$ is the Levi-Civita covariant derivative of\footnote{In earlier works \cite{Gibbons_Werner_2008,Ono_GM_lensing_2017,HallaPerlick2020}
on gravitational lensing from the Gauss-Bonnet theorem, to the same
underlying quotient manifold $\Sigma$, a different Riemannian metric
$\gamma_{ij}=e^{-2\Phi}h_{ij}$ (``generalized optical'' metric
\cite{Ono_GM_lensing_2017}) is associated. The motivation being that,
in the static case $\mathcal{A}_{i}=0$, $\gamma_{ij}$ yields an
optical metric \cite{Gibbons_Werner_2008} (``Fermat'' metric \cite{HallaPerlick2020}),
in the sense that the projection $C=\pi_{\Sigma}C^{(4)}$ of null
geodesics $C^{(4)}$ onto $\Sigma$ yields geodesics with respect
to the metric $\gamma_{ij}$. The two approaches are equivalent, working
with $h_{ij}$ being more suitable for our purposes, by making explicit
the role of the inertial fields $\vec{G}$ and $\vec{H}$.} $(\Sigma,h)$, with Christoffel symbols \eqref{eq:ChristoffelSpace},
as defined above. For light rays, $\dot{C}=\vec{k}$, $\|\dot{C}\|^{2}=h_{\alpha\beta}k^{\alpha}k^{\beta}=\nu^{2}$
{[}cf. Eq. \eqref{eq:vrelLight}{]}, and $e_{\hat{1}}=\vec{v}$. Recalling
that the left-hand member of (\ref{eq:QMGeoGeneral}) is the 3D covariant
acceleration $\tilde{\nabla}_{\dot{C}}\dot{C}$, by Eq. \eqref{eq:Geolightv}
we have $\tilde{\nabla}_{\dot{C}}\dot{C}=\nu^{2}[\vec{G}+\vec{v}\times\vec{H}]$;
hence,
\begin{equation}
\kappa_{{\rm g}}=\langle\vec{G},e_{\hat{2}}\rangle+\langle\vec{v}\times\vec{H},e_{\hat{2}}\rangle\equiv G^{\hat{2}}+(\vec{v}\times\vec{H})^{\hat{2}}\ .\label{eq:GeoCurvatureExplicit}
\end{equation}
For the setting in Fig. \ref{fig:EinsteinRing}, in the equatorial
plane, $(\vec{v}_{\pm}\times\vec{H})^{\hat{2}}=-\|\vec{H}\|$; hence,
the gravitomagnetic contributions to the convergence angle $\theta_{{\rm R}}$
in Eq. \eqref{eq:ConvergenceGaussBonnet} have \emph{opposite} signs.
In particular, for the case where the rays are approximately symmetric,
as required for a nearly perfect Einstein ring, we have, due to the
reflection symmetry about the source-lens-observer axis, $\int_{C_{+}}\|\vec{H}\|d\lambda\approx\int_{C_{-}}\|\vec{H}\|d\lambda$
and $\int_{C_{+}}G^{\hat{2}}d\lambda\approx-\int_{C_{-}}G^{\hat{2}}d\lambda$;
therefore,
\[
\int_{C_{+}}\kappa_{{\rm g}}d\lambda-\int_{C_{-}}\kappa_{{\rm g}}d\lambda\approx2\int_{C_{+}}G^{\hat{2}}d\lambda\ ,
\]
so the gravitomagnetic force does not contribute to $\theta_{{\rm R}}$
in Eq. \eqref{eq:ConvergenceGaussBonnet}. Indeed, the effect of the
gravitomagnetic force would be to cause the light rays that reach
an observer aligned with the source and foreground galaxy to arrive
at different angles from each side, as exemplified in Appendix \ref{sec:LensingSchwKerr}
for the case of a Kerr black hole. Moreover, in the far field regime,
the gravitomagnetic field of a galaxy (as for any spinning body) is
dipole-like, as illustrated in Fig. \ref{fig:DipoleDeflection}; hence,
light rays with impact parameter $\vec{b}$ orthogonal to the equatorial
plane are deflected in a direction parallel to that plane (thus orthogonal
to $\vec{b}$), which again produces no convergence, cf. Secs. \ref{subsec:Lensing-around-a}
and Appendix \ref{sec:LensingSchwKerr}.
 As a consequence, not only $\vec{H}$ does not contribute, as it
acts against the formation of Einstein rings (actually precluding
them for small light sources). This is well known in the case of the
Kerr spacetime \cite{Perlick_Lensing_2004,RauchBlandford_Lensing_Kerr,Bozza_Kerr_Equat,Sereno_De_Luca_Kerr},
discussed in Appendix \ref{sec:LensingSchwKerr}, and for generic
rotating sources in the weak field limit \cite{Sereno_Weak}, discussed
in Sec. \ref{subsec:Lensing-around-a} below. This is in spite of,
for astrophysical lenses, $\vec{H}$ (thus $\vec{v}\times\vec{H}$)
being smaller (typically much smaller) than $\vec{G}$ along the light
ray's path. 

Observe now the crucial difference in magnitude (for a given $\vec{H}$)
between the gravitomagnetic forces $\vec{v}\times\vec{H}$ on light
and on massive astrophysical bodies, 
\begin{align*}
{\rm photons} & :\ v=1\ \Rightarrow\ |\vec{v}\times\vec{H}|\sim|\vec{H}|\\
{\rm stars\ in\ galaxy} & :\ v\apprle10^{-3}\ \Rightarrow\ |\vec{v}\times\vec{H}|\apprle10^{-3}|\vec{H}|\ .
\end{align*}
By Eq. \eqref{eq:QMGeov}, in order for $\vec{v}\times\vec{H}$ to
have a relevant impact on the galactic rotation curves, one would
need $|\vec{H}|\gtrsim10^{3}|\vec{G}|$. Such a gravitomagnetic field
(impossible for a physical source, cf. Sec. \ref{subsec:Lensing-around-a})
would yield a gravitational lens with a gravitomagnetic force three
orders of magnitude larger than the gravitoelectric term: $|\vec{v}\times\vec{H}|\sim|\vec{H}|\gtrsim10^{3}|\vec{G}|$
(thus than the Newtonian contribution). This, besides precluding Einstein
rings, would produce bending angles much larger than observed, as
can be checked using the Gauss-Bonnet theorem (employing it, with
$\kappa_{{\rm g}}$ as given in Eq. \eqref{eq:GeoCurvatureExplicit}
above, to the quadrilateral in e.g. Fig. 2 of \cite{Ono_GM_lensing_2017};
cf. also Eq. (30) therein), or, in linearized theory, using Eq. \eqref{eq:GMdeflection}
below, leading to extreme lensing effects also very different from
those observed. This is a trait common to the ``galactic'' models
based on gravitomagnetism proposed in the literature, as exemplified
in Secs. \ref{subsec:Lensing-around-a} and \ref{subsec:The-Balasin-Grumiller-(BG)}
below.

\section{Nonlinear GR effects work \emph{against} attraction\label{subsec:Non-linear-GR-effects}}

The analysis in the previous section shows that not only the gravitomagnetic
field $\vec{H}$ cannot replace the role of dark matter in gravitational
lensing, as the observed deflection angles constrain $\vec{H}$ in
galaxies to be, at most, within the order of magnitude of $\vec{G}$
(actually having to be considerably smaller, in order to allow for
the Einstein rings). For the motion of stars, since $v\apprle10^{-3}$,
this effectively renders the gravitomagnetic term $\vec{v}\times\vec{H}$
in Eq. \eqref{eq:QMGeov} irrelevant compared to the gravitoelectric
term. The measured star velocities constrain likewise the magnitude
of $\vec{G}$ to $|\vec{G}|r\sim v^{2}\apprle10^{-6}$; from \eqref{eq:vgeo},
we have thus, for the velocity of the circular equatorial geodesics,
\begin{equation}
\|v_{{\rm geo\pm}}^{\alpha}\|=\sqrt{r_{{\rm w}}G_{r_{{\rm W}}}}+O(10^{-9};10^{-6}|\vec{H}|/|\vec{G}|)\ ,\label{eq:vgeoGE}
\end{equation}
which is ruled by $G^{\alpha}$, as expected (and formally similar
to the Keplerian law). We stress that the approximate expression \eqref{eq:vgeoGE}
follows merely from observational constrains, not linearized theory;
indeed $\vec{G}$, so far, remains a nonlinear field. The same conclusion\footnote{As discussed in Sec. III F of \cite{Costa_e_al_Frames}, rotation
curves are a notion of Newtonian and post-Newtonian theories, whose
generalization to exact GR is not unambiguous. The angular velocity
\eqref{eq:OmegaGeo} and the relative velocity \eqref{eq:vgeo} are
two precise notions, both yielding the usual rotation curve in the
Newtonian limit, but which are however distinct (one being an angular
velocity with respect to coordinate time $t$, the other a velocity
with respect to the static observers, hence to their proper time $\tau$,
obeying $d\tau=\sqrt{-g_{00}}dt$).} can be reached from the angular velocity \eqref{eq:OmegaGeo}, which
yields $\Omega_{{\rm geo}\pm}=e^{{\rm \Phi}}(-2G_{r}/g_{\phi\phi,r})^{1/2}(1+[{\rm terms}<10^{-3}])$
(where one can take $g_{\phi\phi}=r^{2}$, since it is also well known
from experiment that the galactic space metric is very weakly curved). 

Thus it remains only to clarify whether nonlinear GR effects can amplify
$\vec{G}$ in order to produce an attractive effect able to sustain
the rotation curves. It is clear however, from the first of Eqs. \eqref{eq:GFieldEq},
that the nonlinear terms ${\vec{G}}^{2}$ and ${\vec{H}}^{2}/2$ act
as effective negative ``energy'' sources for $\vec{G}$ \cite{ZonozBell1998},
\emph{countering} the attractive effect of the source term $2\rho+T_{\ \alpha}^{\alpha}$
($=\rho$ for dust). Hence, they only \emph{aggravate} the need for
dark matter. This is manifest already in post-Newtonian (PN) theory,
see Sec. \ref{subsec:Post-Newtonian-theory} below. It is also clear,
based on PN theory, that the effect is anyway negligible in any realistic
galactic model, given the weakness of the galactic gravitational field
(cf. Secs. \ref{subsec:System-of-} and \ref{subsec:Self-gravitating-disks}).
It is huge, however, in some galactic models proposed in the literature.
Namely, in the Balasin-Grumiller dust model (Sec. \ref{subsec:The-Balasin-Grumiller-(BG)}
below), the effect is such that it completely kills off the attractive
contribution from the dust's mass density $\rho$, yielding $\vec{G}=0$
\cite{Costa_e_al_Frames}.

\section{Relevant examples}

\subsection{Post-Newtonian theory\label{subsec:Post-Newtonian-theory}}

In the post-Newtonian (PN) expansion of general relativity, at first
(1PN) order, the metric can be written, in geometrized units, as \cite{Damour:1990pi,Soffel2009IAU,Kaplan:2009}
\begin{align}
g_{00} & =-1+2w-2w^{2}+O(6)\ ;\nonumber \\
g_{i0} & =\mathcal{A}_{i}+O(5)\ ;\qquad g_{ij}=\delta_{ij}\left(1+2w\right)+O(4)\ ,\label{eq:PNmetric}
\end{align}
where $O(n)\equiv O(\epsilon^{n})$, $\epsilon$ is a small \emph{dimensionless}
parameter such that $U\sim\epsilon^{2}$, $U$ is minus the Newtonian
potential, and $w=U+O(4)$ consists of the sum of the Newtonian potential
$U$ plus \emph{nonlinear} terms of order $\epsilon^{4}$. The bodies'
velocities are assumed such that $v\lesssim\epsilon$ (since, for
bounded orbits, $v\sim\sqrt{U}$), and time derivatives increase the
degree of smallness of a quantity by a factor $\epsilon$; for example,
$\partial U/\partial t\sim Uv\sim\epsilon U$. The geodesic equation
for a test particle can be written as \cite{Damour:1990pi}
\begin{equation}
\frac{d^{2}\vec{x}}{dt^{2}}=(1+v^{2}-2U)\vec{G}+\vec{v}\times\vec{H}-3\frac{\partial U}{\partial t}\vec{v}-4(\vec{G}\cdot\vec{v})\vec{v}+O(6)\ \label{eq:GeoPN}
\end{equation}
with
\begin{equation}
\vec{G}=\nabla w-\frac{\partial\vec{\mathcal{A}}}{\partial t}+O(6)\ ;\quad\vec{H}=\nabla\times\vec{\mathcal{A}}+O(5)\ .\label{eq:GEMfieldsPN}
\end{equation}
In the stationary case (i.e., neglecting all time derivatives), Eq.
\eqref{eq:GeoPN} yields the post-Newtonian limit of Eq. \eqref{eq:QMGeo},
as can be seen noting that $\tilde{D}\vec{U}/d\tau=(U^{0})^{2}d^{2}\vec{x}/dt^{2}-v^{2}\vec{G}+4(\vec{G}\cdot\vec{v})\vec{v}+O(6)$
and $(U^{0})^{-2}=1-v^{2}-2U+O(4)$.

\subsubsection{System of $N$ point bodies\label{subsec:System-of-}}

For a system of $N$ gravitationally interacting point particles,
the metric potentials read, in the \emph{harmonic gauge} (e.g. \cite{Damour:1990pi,SoffelKlioner2008,Kaplan:2009,WillPoissonBook}),
\begin{align}
w & =\sum_{{\rm a}}\frac{M_{{\rm a}}}{r_{{\rm a}}}\left(1+2v_{{\rm a}}^{2}-\sum_{{\rm b\ne a}}\frac{M_{{\rm b}}}{r_{{\rm a}{\rm b}}}-\frac{1}{2}\vec{r}_{{\rm a}}\cdot\vec{a}_{{\rm a}}-\frac{(\vec{r}_{{\rm a}}\cdot\vec{v}_{{\rm a}})^{2}}{2r_{{\rm a}}^{2}}\right)\nonumber \\
\vec{\mathcal{A}} & =-4\sum_{{\rm a}}\frac{M_{{\rm a}}}{r_{{\rm a}}}\vec{v}_{{\rm a}}\ ;\qquad U=\sum_{{\rm a}}\frac{M_{{\rm a}}}{r_{{\rm a}}}\ ,\label{eq:PNPot_Nbodies}
\end{align}
where $M_{{\rm a}}$ is the mass of particle ``a'', $\vec{r}_{{\rm a}}\equiv\vec{x}-\vec{x}_{{\rm a}}$,
$\vec{x}$ is the point of observation, $\vec{x}_{{\rm a}}$ is the
instantaneous position of particle ``${\rm a}$'', $\vec{v}_{{\rm a}}=\partial\vec{x}_{{\rm a}}/\partial t$
its velocity, $\vec{a}_{{\rm a}}=\partial\vec{v}_{{\rm a}}/\partial t$
its \emph{coordinate} acceleration, and $\vec{r}_{{\rm a{\rm b}}}\equiv\vec{x}_{{\rm a}}-\vec{x}_{{\rm b}}$.
The gravitoelectric and gravitomagnetic fields \eqref{eq:GEMfieldsPN}
read
\begin{align}
\vec{G} & =-\sum_{{\rm a}}\frac{M_{{\rm a}}}{r_{{\rm a}}^{3}}\vec{r}_{{\rm a}}\left[1-2\sum_{{\rm b}}\frac{M_{{\rm b}}}{r_{{\rm b}}}-\sum_{{\rm b\ne a}}\frac{M_{{\rm b}}}{r_{{\rm a}{\rm b}}}\right.\nonumber \\
 & \left.+2v_{{\rm a}}^{2}-\frac{3(\vec{r}_{{\rm a}}\cdot\vec{v}_{{\rm a}})^{2}}{2r_{{\rm a}}^{2}}-\frac{1}{2}(\vec{r}_{{\rm a}}\cdot\vec{a}_{{\rm a}})\right]\nonumber \\
 & +3\sum_{{\rm a}}\frac{M_{{\rm a}}}{r_{{\rm a}}^{3}}(\vec{r}_{{\rm a}}\cdot\vec{v}_{{\rm a}})\vec{v}_{{\rm a}}+\frac{7}{2}\sum_{{\rm a}}M_{{\rm a}}\frac{\vec{a}_{{\rm a}}}{r_{{\rm a}}}\ ;\label{eq:GPN}\\
\vec{H} & =-4\frac{M_{{\rm a}}}{r_{{\rm a}}^{3}}\sum_{{\rm a}}\vec{v}_{{\rm a}}\times\vec{r}_{{\rm a}}\ .\label{eq:HPN}
\end{align}

In the case of a single body at rest, the potentials \eqref{eq:PNPot_Nbodies}
reduce to $w=M/r$, $\vec{\mathcal{A}}=0$, and \eqref{eq:PNmetric}
yields the 1PN limit of the Schwarzschild metric in isotropic or harmonic
coordinates (obtained, to the accuracy at hand, from the usual Schwarzschild
coordinates $\{t,\varrho,\theta,\phi\}$ through the substitution
$\varrho=r+M$; cf., e.g., \cite{WillPoissonBook}, pp. 268-270).
Equations \eqref{eq:GPN}-\eqref{eq:HPN} yield in this case $\vec{H}=0$
and 
\[
\vec{G}=-\frac{M}{r^{3}}\left(1-\frac{2M}{r}\right)\vec{r}\ .
\]
This is smaller than the Newtonian (0PN) gravitoelectric field $\vec{G}_{{\rm N}}=-\vec{r}M/r^{3}$,
showing that the nonlinear contribution \emph{decreases} the gravitational
attraction. The angular velocity of circular geodesics, Eq. \eqref{eq:OmegaGeo},
is
\[
\Omega_{{\rm geo\pm}}=\pm\left[\sqrt{\frac{M}{r^{3}}}-\frac{3}{2}\sqrt{\frac{M^{3}}{r^{5}}}\right]\ ,
\]
which, accordingly, is \emph{slower} than the Newtonian angular velocity
$\Omega_{{\rm N}}=\pm M^{1/2}/r^{3/2}$.

Stars in a galaxy can be considered, to a good approximation, point
masses. In the Milky Way their velocities are of the order $v_{{\rm a}}\sim10^{-3}$,
and their coordinate acceleration $a_{{\rm a}}$ of the order $v_{{\rm a}}^{2}/r\sim10^{-6}/r$;
hence, the velocity and acceleration dependent terms in \eqref{eq:GPN}
are smaller by a $10^{6}$ factor compared to the Newtonian term $\vec{G}_{{\rm N}}=-\sum_{{\rm a}}M_{{\rm a}}\vec{r}_{{\rm a}}/r_{{\rm a}}^{3}$.
The same applies to the velocity dependent terms in \eqref{eq:GeoPN},
by Eqs. \eqref{eq:PNPot_Nbodies}--\eqref{eq:HPN}. The remaining
relativistic corrections are the nonlinear terms in the first line
of Eq. \eqref{eq:GPN} (which are negative, thus decreasing the magnitude
of $\vec{G}$), plus the negative term $-2U\vec{G}$ in \eqref{eq:GeoPN};
all of them \emph{decrease} the attraction, thus working against the
sought effect for explaining the galactic rotation curves (the corrections
being anyway negligible, comparing to the Newtonian terms).

\subsubsection{Self-gravitating disks\label{subsec:Self-gravitating-disks} }

The gravitational field of self-gravitating stationary fluids is described,
to first post-Newtonian order, by the metric \eqref{eq:PNmetric}
with $w=U$ \cite{MachMalec2015,Karkowski_et_al_2016}. The relativistic
Euler equations are integrable provided that the angular momentum
per unit mass $j$ depends only on the angular velocity $\Omega$
\cite{MachMalec2015}. A function $j(\Omega)$, appropriate for toroids
(including thin hollow disks, case that could represent a disk galaxy),
is given in Eq. (8) of \cite{MachMalec2015}, leading to the rotation
curve
\begin{equation}
\Omega=\Omega_{{\rm N}}\left[1-\frac{2}{1-\delta}\Omega_{{\rm N}}^{2}r^{2}-\frac{4h_{{\rm N}}}{1-\delta}\right]-\frac{\mathcal{A}_{\phi}}{r^{2}(1-\delta)}\label{eq:DiskPN}
\end{equation}
(cf. Eq. (18) of \cite{MachMalec2015}), where $\Omega_{{\rm N}}$
is the Newtonian result, $\delta\in[-\infty,0]\setminus\{-1\}$ is
a parameter, and $h_{{\rm N}}$ the specific enthalpy at Newtonian
(0PN) accuracy. The nonlinear contribution in the second term, again,
\emph{slows down} the rotation. This contribution, as well as the
third and fourth terms of Eq. \eqref{eq:DiskPN}, are anyway negligible
for the Milky Way: $\Omega_{{\rm N}}^{2}r^{2}\sim10^{-6}\gtrsim h_{{\rm N}}$;
and $\mathcal{A}_{\phi}$, from its definition in e.g. Eqs. (8.4)
of \cite{WillPoissonBook}, is of order $\mathcal{A}_{\phi}/r^{2}\sim10^{-3}U/r\sim10^{-6}\Omega_{{\rm N}}$
(in agreement with the conclusion in \cite{Ciotti:2022inn}). 

\subsubsection{Lensing around a spinning body\label{subsec:Lensing-around-a}}

The gravitational field of any isolated stationary matter distribution
is described, to linear order, by the metric 
\begin{equation}
ds^{2}=(-1+2U)dt^{2}+2\mathcal{A}_{i}dx^{i}+\delta_{ij}\left(1+2U\right)dx^{i}dx^{j}\ ,\label{eq:LinearMetric}
\end{equation}
which follows from linearizing \eqref{eq:PNmetric} (cf. e.g. Eq.
(27.32) of \cite{stephani_RelativityBook}). Consider light rays from
a distant light source, at impact parameter \textbf{$\vec{b}$}, being
scattered by the distribution (see e.g. Fig. 1 of \cite{Ibanez1983}).
Let $\vec{v}$ be a unit vector parallel to the ray's direction {[}i.e.,
the photon's spatial velocity, as defined in Eq. \eqref{eq:vrelLight}{]}.
The change in the ray's direction, $\Delta\vec{v}=\vec{v}_{{\rm f}}-\vec{v}_{{\rm in}}$,
is given by \cite{schneider_Ehlers_Falco_lensing,SchaeferGMLensing}%
{} (cf. also \cite{Ibanez1983}) 
\begin{equation}
\Delta\vec{v}=2\int_{-\infty}^{\infty}\vec{G}dt+\Delta\vec{v}_{{\rm H}}\ ;\qquad\Delta\vec{v}_{{\rm H}}=\int_{-\infty}^{\infty}\vec{v}\times\vec{H}dt\ ,\label{eq:GMdeflection}
\end{equation}
where $\vec{G}=\nabla U$ and $\Delta\vec{v}_{{\rm H}}$ is the gravitomagnetic
deflection. The gravitomagnetic vector potential of the distribution
is $\vec{\mathcal{A}}=2\vec{r}\times\vec{S}/r^{3}$, where $\vec{S}$
is the angular momentum. Notice that this is a dipole-type potential,
of dipole moment $-2\vec{S}$. The corresponding gravitomagnetic field
is {[}cf. Eqs. \eqref{eq:GEMfieldsPN}{]} $\vec{H}=2\vec{S}/r^{3}-6(\vec{S}\cdot\vec{r})\vec{r}/r^{5}$,
formally identical to the magnetic field of a magnetic dipole of moment
$\vec{\mu}=-2\vec{S}$, as is well known (e.g. \cite{CiufoliniWheeler}).
Equations \eqref{eq:GMdeflection} yield in this case \cite{IbanezMartin,Ibanez1983}
\[
\Delta\vec{v}_{{\rm H}}=\frac{4}{b^{2}}\left[\frac{2}{b^{2}}(\vec{b}\cdot(\vec{v}_{{\rm in}}\times\vec{S})\vec{b}-(\vec{v}_{{\rm in}}\times\vec{S})\right]\ .
\]
For pairs of light rays at opposite impact parameters $\pm\vec{b}$
(see Fig. \ref{fig:DipoleDeflection}), $\Delta\vec{v}_{{\rm H}}$
is the same, thus not contributing to produce convergence. For $\vec{v}_{{\rm in}}$
and $\vec{b}$ mutually orthogonal and lying in the equatorial plane
(e.g. $\vec{v}_{{\rm in}}=\|\vec{v}_{{\rm in}}\|\vec{e}_{x}$, $\vec{b}=\pm\|\vec{b}\|\vec{e}_{y}$,
$\vec{S}=\|\vec{S}\|\vec{e}_{z}$, as depicted in Fig. \ref{fig:DipoleDeflection}),
the photons are deflected in the direction $\hat{v}_{{\rm in}}\times\hat{S}$
($=-\vec{e}_{y}$, in Fig. \ref{fig:DipoleDeflection}) along the
equatorial plane; for the same $\vec{v}_{{\rm in}}$, but now $\vec{b}$
orthogonal to the equatorial plane ($\vec{b}\parallel\pm\vec{S}$),
the photons are again deflected parallel to the equatorial plane,
but now in the opposite direction $-\hat{v}_{{\rm in}}\times\hat{S}$.

Considering the Newtonian potential of a spherical source, $U=M/r$,
and $0\le S/M^{2}<1$, Eq. \eqref{eq:LinearMetric} yields the linearized
Kerr metric. It produces weak lensing images very similar to those
displayed for the exact metric in Figs. \eqref{fig:BHImages}(c)--(d).
Namely, for large enough light sources (see Appendix \ref{sec:LensingSchwKerr}),
the lens acts like a shifted non-spinning lens \cite{Sereno_Luca_2006,Asada_rotating_lens}.
For extended objects (like fast-spinning stars) larger values of $S/M^{2}$
are possible; but still the gravitomagnetic field $\vec{H}$ along
the ray's trajectory is smaller than (or, at best, when grazing the
object, within the order of magnitude of) $\vec{G}$, since $S\lesssim Mv_{{\rm rot}}R$
(where $R$ is the body's radius and $v_{{\rm rot}}$ its rotational
velocity), and so in the body's exterior $|\vec{H}|/|\vec{G}|\sim v_{{\rm rot}}R/r<1$,
with $R/r<1$ and $v_{{\rm rot}}<1$. 

To entertain the possibility of $\vec{H}$ having an impact on the
galactic rotation curves, however, a much larger field would need
to be considered, as discussed in Sec. \ref{subsec:Lensing}. According
to Eq. \eqref{eq:QMGeov}, one would need $|\vec{H}|\gtrsim10^{3}|\vec{G}|$
since, for stars in a galaxy, $v\apprle10^{-3}$. Besides impossible
for a rotating body, as shown above, such $\vec{H}$ would, by Eq.
\eqref{eq:GMdeflection} (since, for light, $v=1$) lead to extreme
deflection angles, orders of magnitude larger than observed, and of
a very different type. This can be exemplified with the gravitomagnetic
equatorial potential $\mathcal{A}_{\phi}$ in Eq. (23) of \cite{LasenbyHobsonBarker},
claimed in some literature to flatten the rotation curves (but shown
in \cite{LasenbyHobsonBarker} to be actually sourced by unphysical
singularities). We have, using the values in \cite{LasenbyHobsonBarker}
for the galaxy NGC 1560, and the Newtonian potential $U$ in Eq. (15)
therein, $|\vec{H}|/|\vec{G}|>Kr^{-7.3/30}$, with $K=6400\,{\rm kpc}^{7.3/30}=1/C$,
thus $|\vec{H}|/|\vec{G}|>10^{3}$ throughout the whole galaxy ($R\lesssim10\,{\rm kpc}$).
Numerical computation of the ray trajectories yields an equatorial
plot resembling that in Fig. \ref{fig:BGLensing}, with rays hugely
deflected in the same direction on both sides of the lens, and not
even crossing along the lens-source axis.

\subsection{The Balasin-Grumiller (BG) \textquotedblleft galactic\textquotedblright{}
model\label{subsec:The-Balasin-Grumiller-(BG)}}

The BG solution is described by the metric \cite{BG} 
\begin{figure*}
\includegraphics[width=0.3\paperheight]{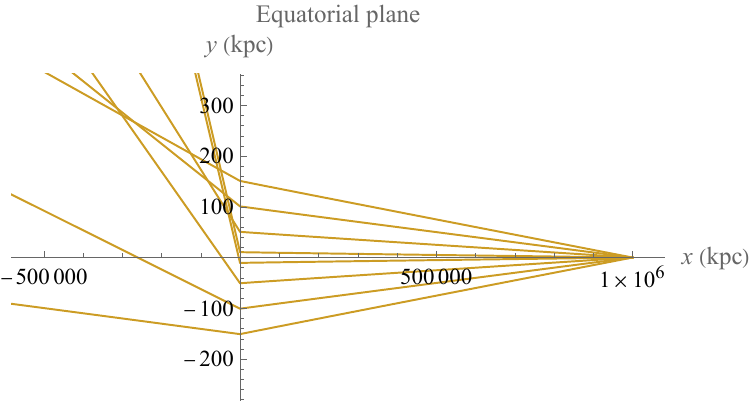}~~~~~~~\includegraphics[width=0.26\paperheight]{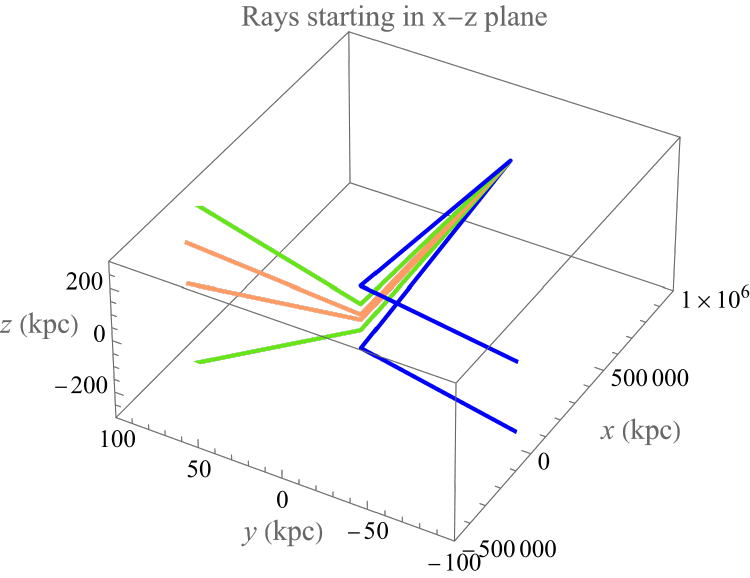}\caption{\label{fig:BGLensing}Gravitational light deflection (numerical results)
in the Balasin-Grumiller solution for $r_{0}=1\,{\rm kpc}$, $R=100\,{\rm kpc}$,
$V_{0}=220\,{\rm km/s}$, $\mu\approx1$, and a light source at $(x,y,z)=(10^{3}{\rm Mpc},0,0)$.
Left panel: in the equatorial plane, light rays through opposite sides
of the body are deflected in the same direction. Right panel: light
rays starting out in the $z_{O}x$ plane. Each color corresponds to
a pair of rays with symmetric angles about the equator; these are
made to diverge along the $z$-direction, and are moreover deflected
along the positive $y$-direction if they approach the $z$-axis at
$|z|<R$, and in the opposite direction at $|z|>R$. This is consistent
with the presence of a pair of axial (oppositely charged) NUT rods
located at $r_{0}<|z|<R$, whose gravitomagnetic field (Fig. 5 of
\cite{Costa_e_al_Frames}) drives the deflection. Multiple images
of the source occur at some points {[}see Fig. \ref{fig:Images-BG}(d){]},
namely on the $y>0$ side of the equatorial plane, where light rays
cross; but no Einstein rings are possible.}
\end{figure*}
\begin{figure*}
\includegraphics[width=1\textwidth]{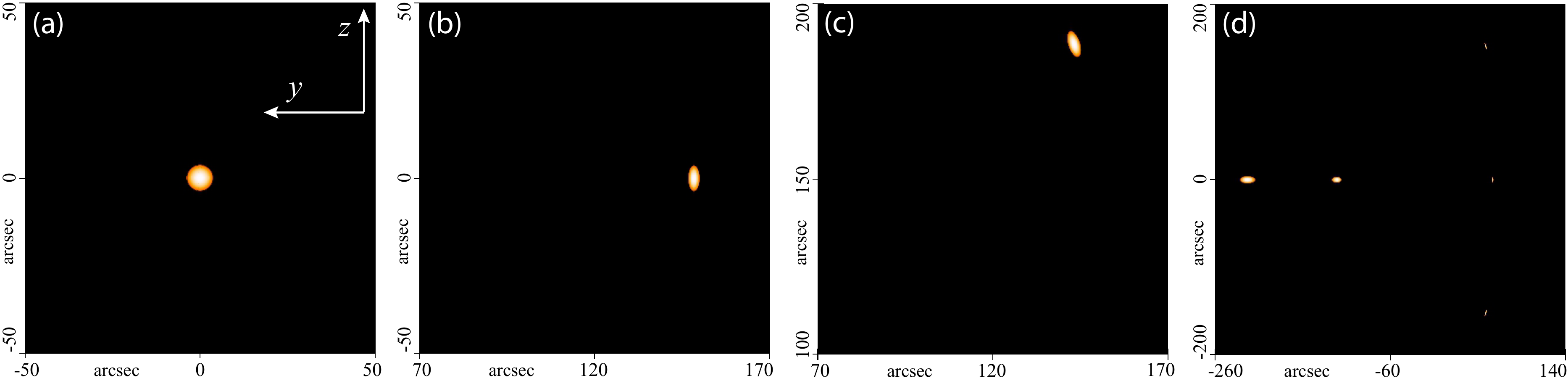}

\caption{\label{fig:Images-BG}Images corresponding to the setting in Fig.
\ref{fig:BGLensing}, for a spherical light source of of radius $R_{{\rm s}}=20\,{\rm kpc}$,
at a distance $1.1\times10^{3}\,{\rm Mpc}$ from the observer, located
along the $x$-axis at $(x_{\mathcal{O}},y_{\mathcal{O}},z_{\mathcal{O}})=(-10^{2}{\rm Mpc},0,0$).
Angular coordinates represent altitude (zero in the equatorial plane)
and azimuth (zero along the negative $x$-semi-axis) in the observer's
sky. Panel (a) is the unlensed image of the light source, located
at $(x_{{\rm s}},y_{{\rm s}},z_{{\rm s}})=(10^{3}\text{Mpc},0,0$),
in Minkowski spacetime. Panels (b)-(c) are images in the Balasin-Grumiller
spacetime. In (b) the source is again located at $(x_{{\rm s}},y_{{\rm s}},z_{{\rm s}})=(10^{3}{\rm Mpc},0,0$)
(i.e., observer, lens, and source all aligned along the $x$-axis);
the result is a single image, oblately deformed, and shifted in the
negative $y$-direction. Panel (c) is the single image of a source
located, in the $x_{O}z$ plane, at $(x_{{\rm s}},y_{{\rm s}},z_{{\rm s}})=$$(10^{3}{\rm Mpc},0,1\,{\rm Mpc})$.
In panel (d) the source is located in the equatorial plane at $(x_{{\rm s}},y_{{\rm s}},z_{{\rm s}})=$$(10^{3}{\rm Mpc},1.64\,{\rm Mpc},0)$
{[}i.e., $(r_{{\rm s}},\phi_{{\rm s}},z_{{\rm s}})=$$(10^{3}{\rm Mpc},339\,{\rm arcsec},0)${]}.
The setting is equivalent to an observer at $(x_{\mathcal{O}},y_{\mathcal{O}},z_{\mathcal{O}})=(-10^{5}{\rm kpc},164\,{\rm kpc},0$)
in the left panel of Fig. \ref{fig:BGLensing}; the result are multiple
distorted images of the source. All the images correspond to numerical
simulations performed with the GYOTO ray-tracing code \cite{Gyoto}.}
 
\end{figure*}
\begin{align}
ds^{2} & =-(dt-Nd\phi)^{2}+r^{2}d\phi^{2}+e^{\mu}(dr^{2}+dz^{2})\ ;\label{eq:BGmetric}\\
N(r,z) & =\frac{V_{0}}{2}\sum_{\pm}\left[\sqrt{(z\pm r_{0})^{2}+r^{2}}-\sqrt{(z\pm R)^{2}+r^{2}}\right]\nonumber \\
 & \quad+V_{0}(R-r_{0})\ ,\label{eq:N}
\end{align}
claimed to describe a galactic dust model in comoving coordinates,
with $r_{0}$ the radius of the bulge region, and $R$ roughly the
radius \cite{Crosta2018} of the galactic disk in the equatorial plane.
It has, however, been recently shown in \cite{Costa_e_al_Frames}
to actually consist of a dust static with respect to the asymptotic
inertial frame (hence totally unsuitable as a galactic model), held
in place by unphysical singularities along the symmetry axis, and
the claimed flat rotation curves to be but an artifact of an unsuitable
choice of reference observers (the zero angular momentum observers---ZAMOs,
which undergo circular motions with respect to inertial frames at
infinity, due to the artificially large frame-dragging effects created
by the singularities). It turns out to be also an archetypal example
for Secs. \ref{subsec:Lensing} and \ref{subsec:Non-linear-GR-effects}. 

Since $\vec{G}=0$ and $\vec{J}=0$ \cite{Costa_e_al_Frames}, Eqs.
\eqref{eq:GFieldEq}--\eqref{eq:HFieldEq} yield $\tilde{\nabla}\cdot\vec{H}=0$,
$\tilde{\nabla}\times\vec{H}=0$, and
\begin{equation}
\tilde{\nabla}\cdot\vec{G}=-4\pi\rho+\frac{1}{2}{\vec{H}}^{2}=0\ .\label{eq:DivGBG}
\end{equation}
Linearization of this equation leads to the empty space equation $\rho=0$;
hence, the solution has no linear or Newtonian limit, being thus a
purely nonlinear solution. In fact, it is an extreme example of the
phenomenon discussed in Sec. \ref{subsec:Non-linear-GR-effects}:
the repulsive action of the nonlinear contribution $\vec{H}^{2}/2$
is such that it \emph{cancels out} exactly the attractive gravitational
effect of the dust's energy density (allowing the gravitoelectric
field $\vec{G}$ to vanish \cite{Costa_e_al_Frames}).

The equation for null geodesics, from Eqs. \eqref{eq:QMGeoGeneral},
\eqref{eq:Geolight}, and \eqref{eq:Geolightv}, becomes here
\[
\frac{d^{2}x^{i}}{d\lambda^{2}}+\Gamma(h)_{jl}^{i}k^{j}k^{l}=\nu\vec{k}\times\vec{H}=\nu^{2}\vec{v}\times\vec{H}\ .
\]
If one approximates, as done in \cite{BG}, $\mu\approx{\rm const}.$,
the space metric $h_{ij}$ becomes flat, $\Gamma(h)_{jl}^{i}$ become
the Christoffel symbols of a cylindrical coordinate system in Euclidean
space, and the spacetime an example of a solution where light deflection
is solely driven by the gravitomagnetic ``force'' $\nu^{2}\vec{v}\times\vec{H}$.
The numerical results, for the parameters suggested in \cite{BG}
for the Milky Way ($r_{0}=1\,{\rm kpc}$, $R=100\,{\rm kpc}$, $V_{0}=220\,{\rm km/s}$)
and a point light source in the equatorial plane at a distance $1\,{\rm Gpc}$
from the origin (within the typical order of magnitude of the sources'
distance of observed Einstein rings, e.g. \cite{Lagattuta_Einstein_ring,Schuldt_et_al_horseshoe_2019}),
are plotted in Fig. \ref{fig:BGLensing}. For a finite light source
of radius $20\,{\rm kpc}$, the resulting images, obtained with the
GYOTO \cite{Gyoto} ray-tracing code, are shown in Fig. \ref{fig:Images-BG}
for different source positions. 

They are consistent with the expectation for the gravitomagnetic field
as plotted in Fig. 5 of \cite{Costa_e_al_Frames}, generated by a
pair of rods with opposite Newman--Unti--Tamburino (NUT) charges
located along the $z$-axis at $r_{0}<z<R$ and $-R<z<-r_{0}$. In
the equatorial plane, $\vec{H}|_{z=0}=|H|\partial_{z}$ points always
in the same direction orthogonal to the plane; hence, light rays passing
through opposite sides of the galaxy are deflected in the same direction
(the $\partial_{y}$ direction). The rays do not cross along the line
of sight of an observer aligned with the light source and foreground
galaxy, resulting in a single image, shown in Fig. \ref{fig:Images-BG}(b)
for an observer at $x_{\mathcal{O}}=-0.1\,{\rm Gpc}$; such image
is moreover deformed and shifted in the negative $y$-direction (by
a huge angle $\approx150\,{\rm arcsec}$). Observe that, for a spherical
lens with $M=10^{12}M_{\odot}$ (approximately the Milky Way's mass),
this aligned setting would yield instead an Einstein ring with angular
radius \cite{Pinochet_Ring,Petters_2001singularity,Virbhadra_Ellis_Lensing1999}
$\theta_{{\rm E}}=\sqrt{4Md_{{\rm ls}}/(d_{{\rm s}}d_{{\rm l}})}\approx9\,{\rm arcsec}$,
where $d_{{\rm ls}}=x_{{\rm s}}=1\,{\rm Gpc}$ is the distance between
the lens and the source (located at $x_{{\rm l}}=0$), and $d_{{\rm s}}=(x_{{\rm s}}-x_{\mathcal{O}})=1.1\,{\rm Gpc}$
and $d_{{\rm l}}=-x_{\mathcal{O}}=0.1\,{\rm Gpc}$ are the distances
between the observer and, respectively, the source and lens. 

Light sources shifted from the optical axis in the negative $y$-direction
yield also a single image, consistent with the fact that in the $y<0$
side of the setting in Fig. 3 light rays do not cross. The same applies
to sources shifted along the $x_{O}z$ plane, as shown in Fig. \ref{fig:Images-BG}(c).
Some light rays do cross on the $y>0$ side of the setting in Fig.
3, leading to multiple images of the source, shown in Fig. \ref{fig:Images-BG}(d)
(for the equivalent situation of an observer along the $x$-axis and
a source displaced in the positive $y$-direction by the same angle
relative to the axis). These images display moreover a huge angular
separation {[}about $290\,{\rm arcsec}$ in Fig. \ref{fig:Images-BG}(d){]}.
Therefore, it is clear that the BG model cannot produce Einstein rings
such as those observed in \cite{King_first_Einstein_Ring,Lagattuta_Einstein_ring}.
The lensing effects it generates, besides of a very different type,
are moreover of a magnitude much larger than observed.

\subsection{Gravitomagnetic dipole model\label{subsec:Gravitomagnetic-dipole-model}}

\begin{figure*}
\includegraphics[width=0.3\paperheight]{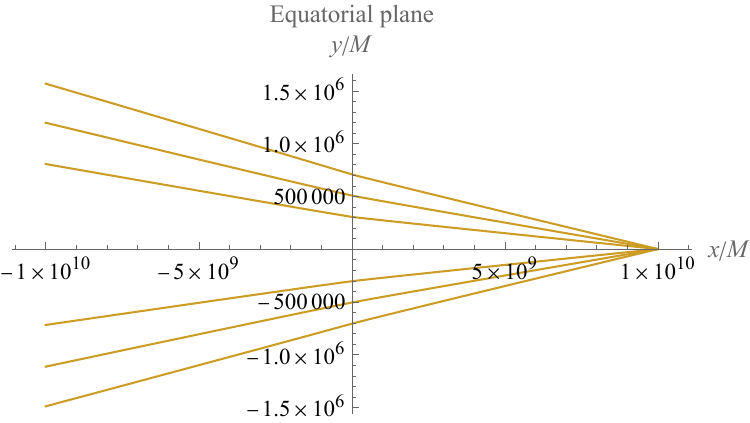}~~~~~~~\includegraphics[width=0.26\paperheight]{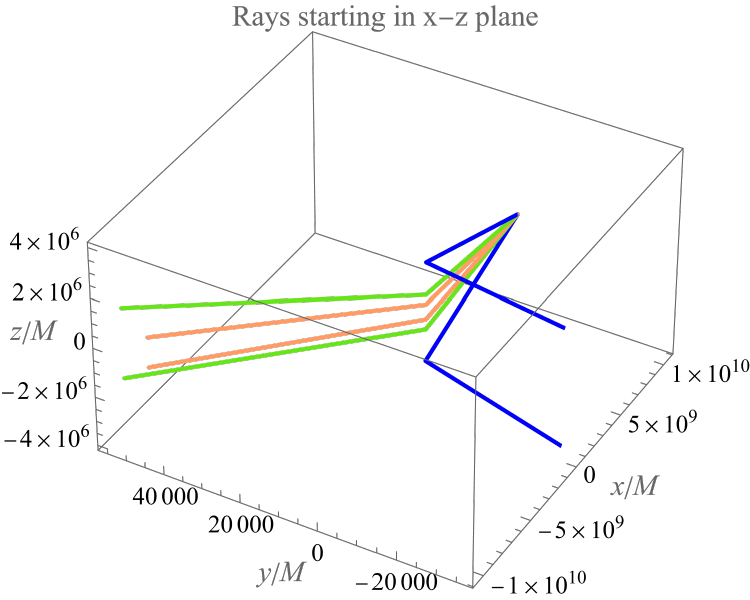}\caption{\label{fig:GMDipoleLensing}Gravitational light deflection (numerical
results) in the gravitomagnetic dipole model, for $Q_{{\rm NUT}}=\pm M$,
separation $2k=3.6\times10^{6}M$, and a light source at a distance
$r=10^{10}M$. No convergence of light rays is produced, which immediately
rules out this setting as a viable galactic model. Left panel: rays
in the equatorial plane are made to diverge. Right panel: light rays
starting out in the $z_{O}x$ plane (each color corresponds to a pair
of rays with symmetric angles about the $x$-axis). They are deflected
orthogonally to that plane; along the positive $y$-direction if they
approach the $z$-axis in the region between the two NUT black holes,
$|z|<k$, and in the opposite direction otherwise, as expected from
the gravitomagnetic field plotted in Fig. \ref{fig:GMDipoleGH}. }
\end{figure*}
\begin{figure*}
\includegraphics[width=0.24\paperheight]{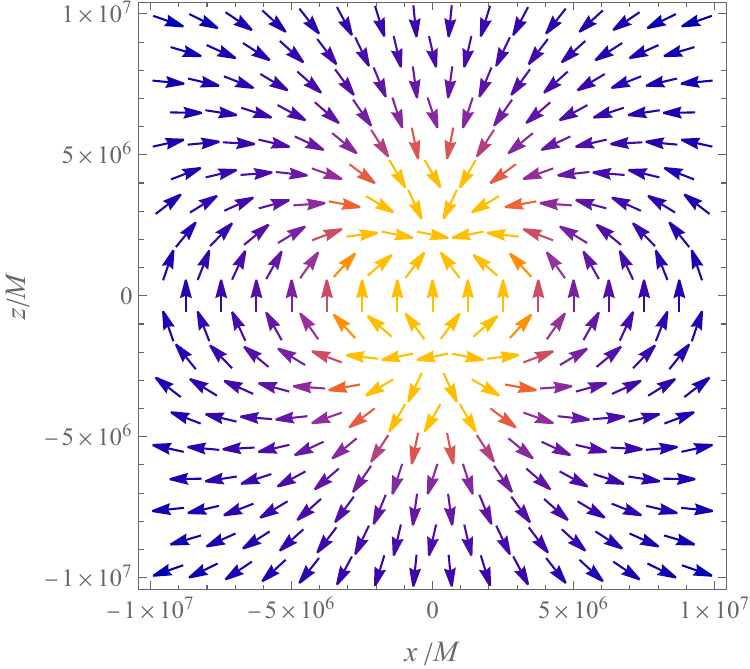}~~~~~~~~~~~~~~~~\includegraphics[width=0.23\paperheight]{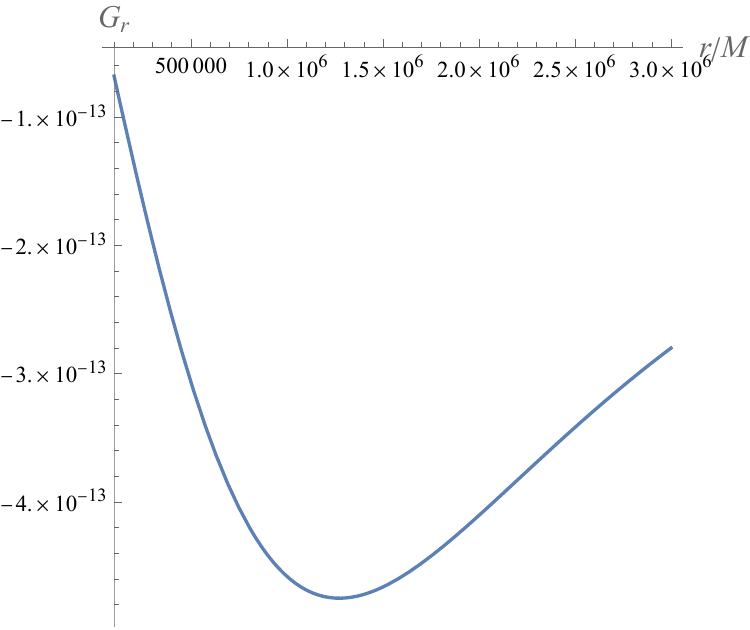}\caption{\label{fig:GMDipoleGH}Left panel: $x$-$z$ plot of the gravitomagnetic
field $\vec{H}$ for the gravitomagnetic dipole model with $Q_{{\rm NUT}}=\pm M$,
and separation $2k=3.6\times10^{6}M$; color represents field strength
(blue=weak, yellow=strong). Right panel: radial component of the equatorial
gravitoelectric field $\vec{G}$, as a function of $r$. Albeit attractive,
$|G_{r}|$ does not monotonically decrease with $r$.}
\end{figure*}
 In \cite{GovaertsGMDipole} a galactic model based on a so-called
``gravitomagnetic dipole'' is proposed. It consists of the solution
obtained in \cite{Clement_GM_Dipole,Manko_et_al_Stringy_NUT} from
a (nonlinear) superposition of two NUT solutions, interpreted as a
pair of NUT ``objects'' of equal masses and equal in magnitude,
but opposite, NUT charges. The lengthy expression for the metric is
given in Eqs. (1)--(10) of \cite{GovaertsGMDipole}. It depends only
on three parameters: the mass $M$ and NUT charge $\pm Q_{{\rm NUT}}$
($\pm\nu$, in the notation of \cite{GovaertsGMDipole,Clement_GM_Dipole})
of each object, and their separation $2k$. For sufficiently large
$k$, the objects are interpreted as a pair of NUT black holes connected
by a spinning Misner string \cite{GovaertsGMDipole}. The gravitomagnetic
field \eqref{eq:GEM1forms} plotted in Fig. \ref{fig:GMDipoleGH}
is indeed consistent with a pair of opposite gravitomagnetic monopoles
(NUT singularities). We choose the values $\nu\simeq M$ and $k=1.8\times10^{6}M$
as proposed in Eqs. (82) and (85) of \cite{GovaertsGMDipole}, claimed
therein\footnote{In \cite{GovaertsGMDipole} a flat velocity curve with $v\approx8\times10^{-4}$
is obtained within the range $4\times10^{5}<r/M<1.5\times10^{6}$,
corresponding, for $M=2.6\times10^{11}M_{\odot}$, to $5\ {\rm kpc}<r<19\ {\rm kpc}$.
This is however actually an artifact of an incorrect approximation
made in Eq. (65) of \cite{GovaertsGMDipole}. An exact computation
using Eqs. \eqref{eq:OmegaGeo} or \eqref{eq:vgeo} reveals a curve
displaying a much less pronounced flattened region, and in a different
range, $1.5\times10^{6}<r/M<3.5\times10^{6}$, where it is almost
indistinguishable from the Newtonian analogue generated by two point
masses separated by a distance $2k$, Eq. (99) of \cite{GovaertsGMDipole}.
A detailed account of the circular geodesics in this metric shall
be given elsewhere.} to yield, for $M=2.6\times10^{11}M_{\odot}$ ($M_{\odot}\equiv$
solar mass), a nearly flat velocity profile for circular equatorial
geodesics approximately matching the observed flat region of the Milky
Way rotation curve. 

The trajectories of light rays emitted from a light source in the
equatorial plane at $r=10^{10}M$, corresponding (again, for $M=2.6\times10^{11}M_{\odot}$)
to $10^{2}{\rm Mpc}$, i.e., roughly 100 times the average distance
between nearby galaxies, is plotted in Fig. \ref{fig:GMDipoleLensing}.
No convergence of light rays is produced (thus no Einstein rings,
nor multiple image of the source); the rays diverge. This completely
rules out this model as a viable galactic model. 

Light rays starting in the equatorial plane diverge along that plane;
the effect, however, is actually not caused by the gravitomagnetic
field. It occurs as well when $Q_{{\rm NUT}}=0\Rightarrow\vec{\mathcal{A}}=\vec{H}=0$
(in which case the setting is held static by the tension of the string).
In fact, the plot corresponding to $Q_{{\rm NUT}}=0$ is only slightly
changed comparing to the left panel of Fig. \ref{fig:GMDipoleLensing},
being then symmetric with respect to the $x$-axis. The effect (which,
in appearance, reminds a repulsive scattering) occurs as well for
massive particles (i.e., timelike worldlines). Inspection of the radial
equation $d^{2}r/d\tau^{2}=-\Gamma_{\alpha\beta}^{r}U^{\alpha}U^{\beta}$
shows that it stems from the weakness of the attractive term $-\Gamma_{00}^{r}E^{2}(g^{00})^{2}=e^{-2\Phi}G^{r}E^{2}$
(see Footnote \ref{fn:Christoffel}; here $E\equiv-U_{\alpha}\partial_{t}^{\alpha}=-U_{0}$
is the particle's energy per unit mass), due to the unusual behavior
of the gravitoelectric field $\vec{G}$, plotted in Fig. \ref{fig:GMDipoleGH}:
although always attractive ($G_{r}<0$), its magnitude does not monotonically
increase with decreasing $r$; in fact, within the region $r/M<1.5\times10^{6}$
(which includes the flat profile range), $|\vec{G}|$ decreases approaching
$r=0$, with $\lim_{r\rightarrow0}\vec{G}=0$. This contrasts with
e.g. the situation in the Schwarzschild geometry (where $|G_{r}|$
monotonically increases until the horizon, where $\lim_{r\rightarrow r_{+}}G_{r}=-\infty$),
or in Levi-Civita or Lewis-Weyl cylinders \cite{Cilindros}, and causes
the centrifugal term $-\Gamma_{\phi\phi}^{r}L^{2}(g^{\phi\phi})^{2}=(1/2)g_{\phi\phi,r}L^{2}(g^{\phi\phi})^{2}/g_{rr}$
to sharply dominate the scattering (hence the apparent repulsion).
Here $L\equiv U_{\alpha}\partial_{\phi}^{\alpha}=U_{\phi}$ is the
test particle's angular momentum per unit mass. The same conclusion
can be reached by inspecting the effective potential $V_{{\rm eff}}(E,L,r)$
that follows from the equation $U^{\alpha}U_{\alpha}=-1\Leftrightarrow(dr/d\tau)^{2}=-2V_{{\rm eff}}$,
so that $d^{2}r/d\tau^{2}=-dV_{{\rm eff}}/dr$. 

Light rays starting out in the $z_{O}x$ plane are deflected orthogonally
to that plane, via the action of the gravitomagnetic field $\vec{H}$.

It should moreover be noted that (as correctly noticed in \cite{GovaertsGMDipole})
the $z$-distance $2k=3.6\times10^{6}M$ between the NUT black holes
is comparable or larger than the galaxy size, which adds to the unrealistic
character of the model.

\section{Conclusion}

We have shown that, in light of the experimentally measured galactic
rotation curves and gravitational lensing, general relativity cannot
resolve the missing mass problem usually attributed to dark matter,
and considering nonlinear corrections only aggravates the problem.
In the process we wrote the equations for light propagation in the
quasi-Maxwell formalism, and by gravitational lensing ruled out any
galactic model (linear or nonlinear) based on gravitomagnetism, as
well as the recent ``dipole'' models (gravitomagnetic or not, linear
or not).

\subsection*{Acknowledgments}

We thank A. Pollo, E. Malec, P. Mach, and D. Grumiller for correspondence
and very useful discussions. We are also indebted to T. Paumard and
F. Vincent for generous and crucial help in implementing the GYOTO
ray-tracing code, and to the anonymous referee for valuable remarks
and suggestions. L.F.C. and J.N. were supported by FCT/Portugal through
Projects UIDB/MAT/04459/2020 and UIDP/MAT/04459/2020.

\appendix

\section{Lensing in the Schwarzschild and Kerr spacetimes\label{sec:LensingSchwKerr}}

\begin{figure*}
\includegraphics[width=0.3\paperheight]{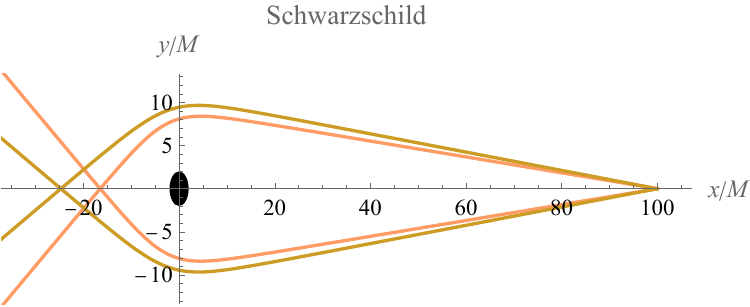}~~~~~~~~~\includegraphics[width=0.3\paperheight]{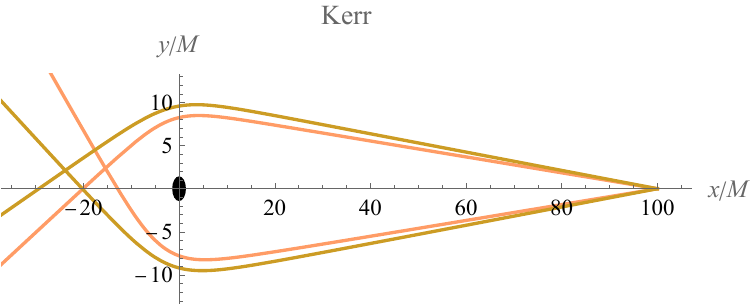}

\includegraphics[width=0.2\paperheight]{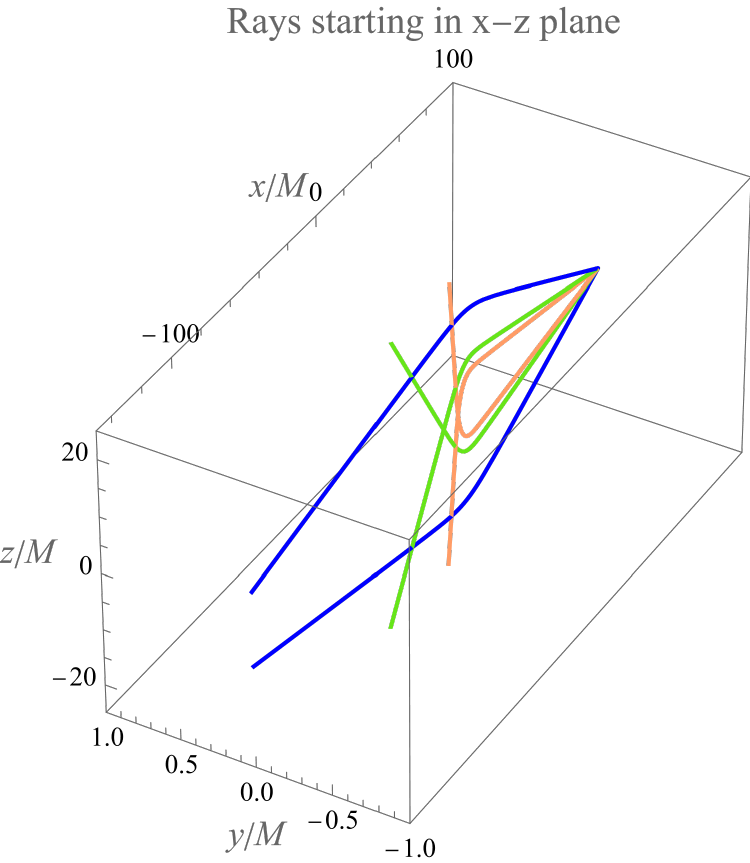}~~~~~~~~~\includegraphics[width=0.27\paperheight]{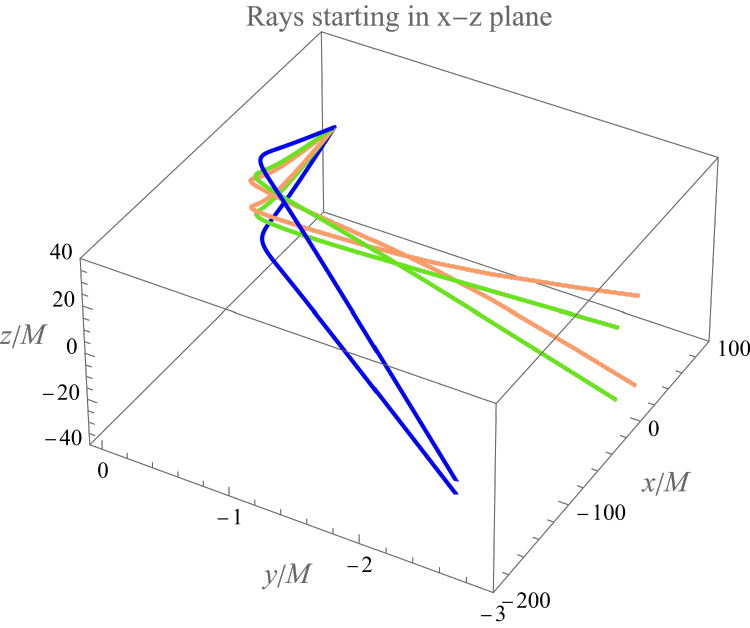}

\caption{\label{fig:KerrSchw}Gravitational light deflection (numerical results)
in the Schwarzschild and in a Kerr spacetime with rotation parameter
$a=S/M=0.9$. The source is at a distance $r=100M$. First row: equatorial
plane; second row: light rays starting out in the $z_{O}x$ plane.
Each color corresponds to a pair of rays with symmetric angles about
the $x$-axis. In the Schwarzschild case, the deflection is the same
for rays starting at equal (in magnitude) angles relative to the $x$-axis
connecting the source and the black hole; this leads to perfectly
circular Einstein rings as viewed by observers along the $x$-axis
{[}Fig. \ref{fig:BHImages} (b){]}. In the Kerr spacetime, the deflection
of equatorial light rays through opposite sides of the black hole
differs; rays starting at equal (in magnitude) angles will not cross
along the $x$-axis; and those that do cross along the $x$-axis arrive
at different angles. Light rays with initial direction along the $z_{O}x$
plane suffer (in addition to the convergence along the $z$-axis),
a deflection along the $y$-direction. These gravitomagnetic effects
do not enhance the convergence of light rays for observers along the
$x$-axis, instead mimicking a non-aligned Schwarzschild lens, as
shown in Figs. \ref{fig:BHImages} (c)-(d).}
\end{figure*}
\begin{figure*}
\includegraphics[width=0.2\paperwidth]{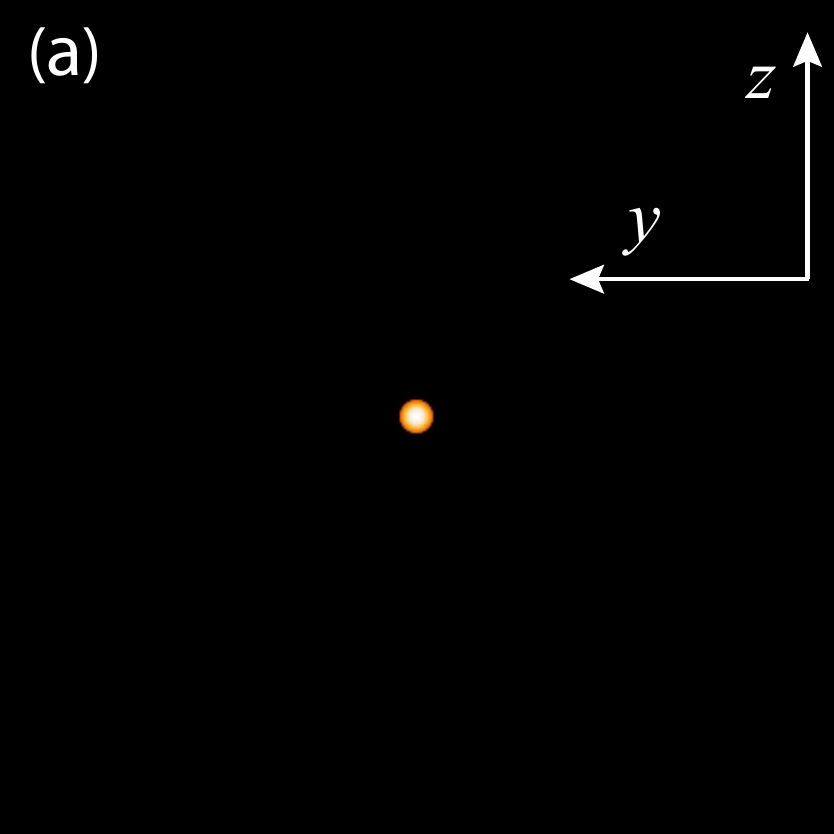}~\includegraphics[width=0.2\paperwidth]{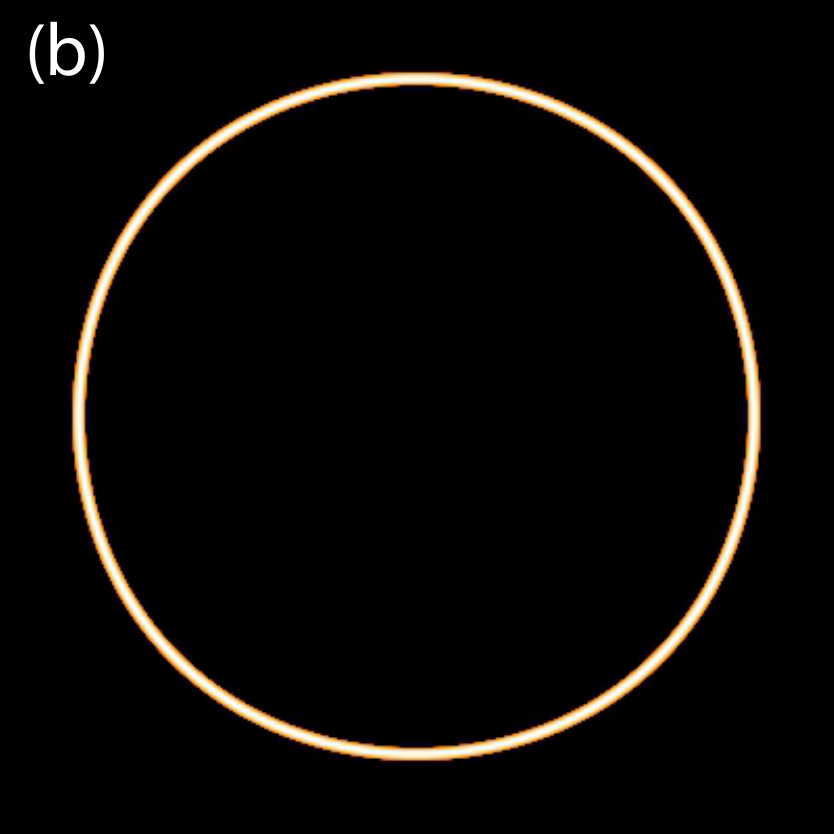}~\includegraphics[width=0.2\paperwidth]{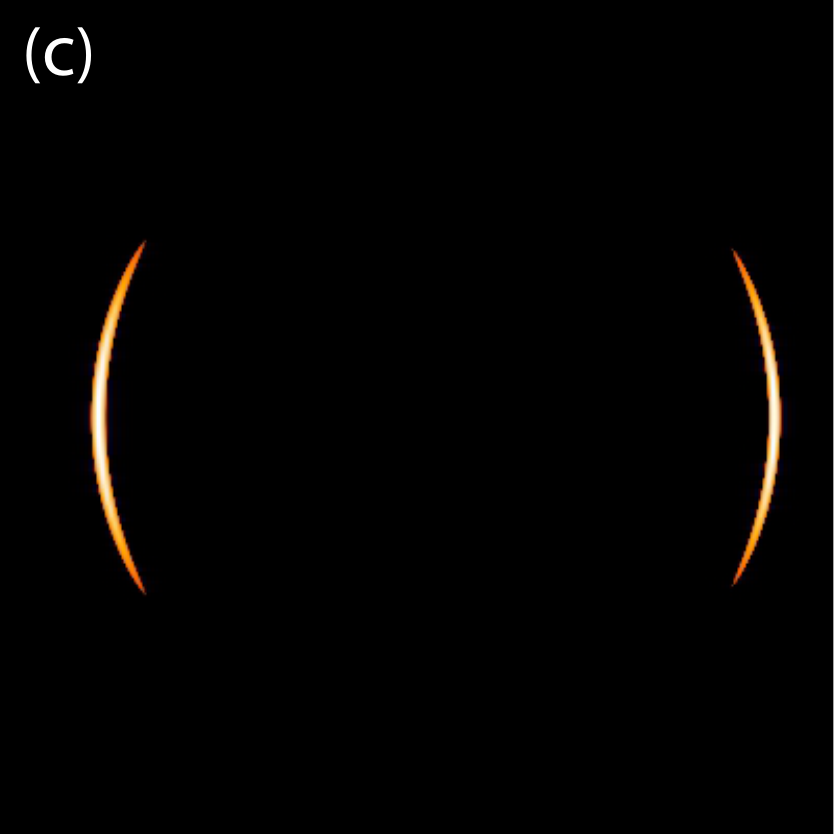}~\includegraphics[width=0.2\paperwidth]{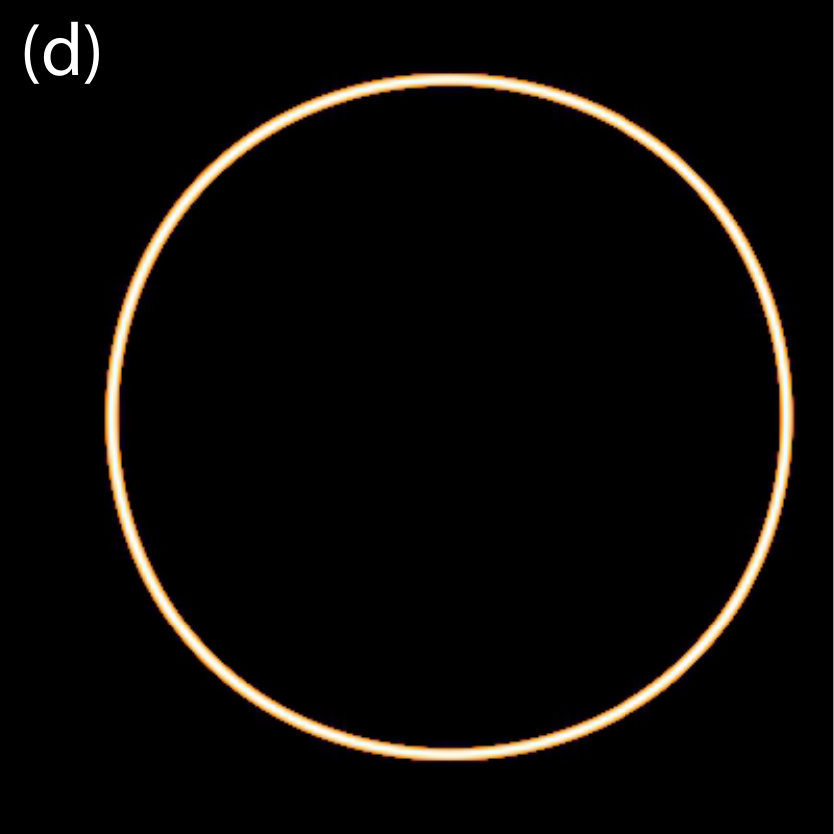}

\caption{\label{fig:BHImages}Images produced by the settings in Fig. \ref{fig:KerrSchw},
for a spherical light source of radius $R_{{\rm s}}=3M$. Panel (a)
is the unlensed image of the light source in Minkowski spacetime.
Panel (b) is the image of the source when placed behind a Schwarzschild
black hole (of mass $M$) along the optical axis (i.e., light source,
black hole, and observer all aligned along the $x$-axis). The source
is located at $(x_{{\rm s}},y_{{\rm s}},z_{{\rm s}})=(100M,0,0)$,
and the observer at $(x_{\mathcal{O}},y_{\mathcal{O}},z_{\mathcal{O}})=(-20M,0,0)$;
the black hole is at the origin. The resulting image is a perfect
Einstein ring. Panel (c): image produced by a similarly aligned setting
for a Kerr black hole with $a=0.9$; it results in a pair of arcs.
Panel (d): for the same Kerr black hole and observer position, the
light source is now placed at the center of the primary caustic $(x_{{\rm s}},y_{{\rm s}},z_{{\rm s}})\approx(100M,-6M,0)$,
which is transversely displaced \cite{Sereno_De_Luca_Kerr} from the
optical axis by $\Delta y\approx-6M$. (The caustic section is therein
about $0.1M$ wide, cf. Eq. (19) of \cite{Sereno_De_Luca_Kerr}, much
smaller than the source's diameter). The image is an Einstein ring
almost identical to panel (b), only transversely shifted in the $y$-direction.
All the images correspond to numerical simulations performed with
the GYOTO ray-tracing code \cite{Gyoto}.}
\end{figure*}
 For comparison with the lensing effects of the models in Secs. \ref{subsec:The-Balasin-Grumiller-(BG)}
and \ref{subsec:Gravitomagnetic-dipole-model}, in order to evince
their unphysical character, we present in Figs. \ref{fig:KerrSchw}
and \ref{fig:BHImages} the corresponding ray trajectories and lensed
images produced by a Schwarzschild black hole (displaying some essential
features common to other compact and spherically symmetric lenses)
and, to display the role of the gravitomagnetic field, of a fast spinning
Kerr black hole. 

In the Schwarzschild case, the deflection is the same for rays starting
at equal (in magnitude) angles relative to the axis connecting the
source and the black hole; this leads to perfect Einstein rings as
viewed by observers along this axis \cite{Perlick_Lensing_2004,LensigInterstellar,Virbhadra_Lensing2008,Virbhadra_Ellis_Lensing1999,KoganTsupko2008}
(``optical axis''; here, the $x$-axis). Namely, the image of an
idealized point source is a (infinitely bright) circle; for a real,
finite source, the image is a ring of finite width \cite{Petters_2001singularity,schneider_Ehlers_Falco_lensing},
exemplified in the simulation in Fig. \ref{fig:BHImages}(b). 

In the Kerr spacetime, the lens is not spherically symmetric; consequently,
for an idealized point source, the Einstein ring does not form \cite{Sereno_De_Luca_Kerr,Perlick_Lensing_2004,RauchBlandford_Lensing_Kerr,Bozza_Kerr_Equat}.
The primary caustic surface \cite{BozzaCausticsKerr} behind the black
hole no longer degenerates into the optical axis, having instead a
finite astroid section (see e.g. Figs. 2--3 in \cite{RauchBlandford_Lensing_Kerr}),
and the image of point sources (inside or outside it) are a finite
set of points (not some 1D contour) \cite{Sereno_De_Luca_Kerr,RauchBlandford_Lensing_Kerr,BozzaCausticsKerr}.
For finite sources, however, rings can still form, when the source
is large enough to cover to whole caustic section \cite{Petters_2001singularity,schneider_Ehlers_Falco_lensing}.
In this case, the impact of the black hole's spin is the following.
The gravitomagnetic field (which is approximately dipole-like, as
depicted in Fig. \ref{fig:DipoleDeflection}) causes the deflection
angles of equatorial light rays through opposite sides of the black
hole to differ (left panel of Fig. \ref{fig:KerrSchw}). Hence, rays
starting at equal but opposite angles relative to the $x$-axis will
not cross along that axis; conversely, those that do, arrive at different
angles. On light rays with initial direction along the $z_{O}x$ plane,
the gravitomagnetic field causes a deflection along the $y$-direction,
causing them not to cross along the $x$-axis (besides not contributing
to their convergence which, as in the Schwarzschild case, is along
the $z$-direction). Consequently, the caustic behind the black hole
is transversely displaced (in the $y$-direction) from the optical
axis, cf. e.g. Fig. 2 in \cite{RauchBlandford_Lensing_Kerr}. As a
result, the image of a finite source along the optical axis will in
general be two arcs, as exemplified in the simulation in Fig. \ref{fig:BHImages}(c);
if the source extends until the caustic location (and covering it)
the arcs will close into a deformed ring, weakened at the poles, and
slightly shifted in the negative $y$-direction. This is similar to
the situation in a Schwarzschild lens of the same mass, but for a
source displaced from the optical axis. For a source centered at the
center of the caustic section, and fully covering it, the image will
be an almost perfect Einstein ring similar to that produced by a Schwarzschild
lens of the same mass for a source along the optical axis, only shifted
along the $y$-direction, as shown in Fig. \ref{fig:BHImages}(d).
The ring's angular diameter (hence the lens power), in particular,
is the same. (Such a ring is also slightly deformed \cite{Bozza_Kerr_Equat};
the effect is however almost unnoticeable in the simulations performed.)
It is thus clear that the gravitomagnetic field does not enhance the
optical effect generating Einstein rings. Indeed, for sources large
enough for the ring to still form, unless their displacement from
the optical axis can be verified by an independent method, one cannot,
based only on the observed Einstein rings, distinguish a spinning
from a non-spinning black hole with the same mass \cite{Sereno_Luca_2006,Asada_rotating_lens}.

\appendix

\bibliographystyle{utphys}
\bibliography{Ref}

\end{document}